\documentclass[preprintnumbers,amsmath,amssymb,showpacs]{revtex4}
\usepackage{epsfig}
\usepackage{color}
\begin{document}
\setlength{\voffset}{1.0cm}
\title{Baryon-baryon scattering in the Gross-Neveu model: the large $N$ solution}
\author{Gerald V. Dunne\footnote{dunne@phys.uconn.edu}}
\affiliation{Department of Physics, University of Connecticut, Storrs, CT 06269}
\author{Christian Fitzner\footnote{fitzner@theorie3.physik.uni-erlangen.de}}
\author{Michael Thies\footnote{thies@theorie3.physik.uni-erlangen.de}}
\affiliation{Institut f\"ur Theoretische Physik III,
Universit\"at Erlangen-N\"urnberg, D-91058 Erlangen, Germany}
\begin{abstract}
The scattering of two Dashen-Hasslacher-Neveu (DHN) baryons in the 
large $N$ Gross-Neveu model is solved exactly using the relativistic, time-dependent Hartree-Fock approach.
Unlike the special case of kink-antikink scattering, the scattering of DHN baryons is sensitive to the back-reaction of the fermions bound inside the baryons. Correspondingly the solution
is much more complicated than the kink-antikink scattering solutions, which can be expressed in terms of sinh-Gordon solitons. Nevertheless, we present a simple ansatz form that gives closed
analytic expressions for both the space-time dependent mean fields 
and the Dirac spinors for all continuum and bound states.
The solution can also be applied to the scattering of polarons and solitons
in conducting polymers. 
\end{abstract}
\pacs{11.10.-z,11.10.Kk}
\maketitle

\section{Introduction}\label{sect1}
\begin{figure}[htb]
\includegraphics[scale=1]{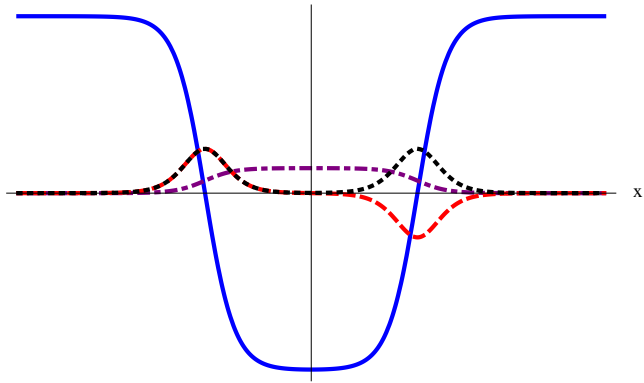}
\caption{A sketch of the anatomy of a DHN baryon, here shown for parameter $y=0.999999$. The scalar field $S_{\rm DHN}(x)$ is shown as the solid blue line going asymptotically to 1
(in units of the dynamically generated fermion mass in the GN model). The fermion density, $\rho=\psi^\dagger\psi$, for the bound valence mode is shown as a black dotted line; note that the 
density is localized at the kink and antikink of the baryon scalar field. The pseudoscalar condensate, $\pi(x)=\bar{\psi}\,i\, \gamma_5 \psi$, for the valence state is plotted as a red dashed 
curve, while the scalar condensate, $\sigma(x)=\bar{\psi}\, \psi$, for the valence state is plotted as a purple 
dot-dashed curve ($\sigma$ has been multiplied by a factor of 100, for visualization purposes). The scalar condensate is nonzero inside the DHN baryon, while the pseudoscalar condensate 
is localized at the edges, on the kink and antikink, like the density $\rho$ but with a change of sign.}
\label{f1}
\end{figure}

The simplest version of the Gross-Neveu (GN) model \cite{L1} is a 1+1 dimensional model quantum field theory of $N$ species of massless, 
self-interacting Dirac fermions with Lagrangian 
\begin{equation}
{\cal L} =  \sum_{k=1}^N \bar{\psi}_k i\partial \!\!\!/ \psi_k + \frac{g^2}{2} \left( \sum_{k=1}^N \bar{\psi}_k \psi_k \right)^2.
\label{a1}
\end{equation}
We consider this model in the 't~Hooft limit, $N \to \infty$, with $Ng^2=$ constant, where semiclassical methods become exact. In spite of its
apparent simplicity, the model (\ref{a1}) gives rise to quite non-trivial physical phenomena of interest for strong interaction particle physics
as well as condensed matter physics. 
This has been established in works spread out over several decades, in an effort which is still ongoing,
as we shall see. The original work from 1974 was motivated in part by the discovery of asymptotic freedom in quantum chromodynamics
(QCD) \cite{L2,L3}, a property which the GN model shares. The main focus in Ref.~\cite{L1} was on spontaneous breaking of the discrete
chiral symmetry $\psi \to \gamma_5 \psi$, fermion mass generation via dimensional transmutation and the scalar $\sigma$ meson. Soon
afterwards, multifermion bound states (baryons) were found by Dashen, Hasslacher and Neveu (DHN) in a study of the associated gap equation \cite{L4}. 
The gap equation solution for the static DHN baryon scalar potential is
\begin{eqnarray}
S_{\rm DHN}(x)=1+y\, \tanh(y\, x-\frac{1}{2} {\rm arctanh}(y))-y\, \tanh(y\, x+\frac{1}{2}  {\rm arctanh}(y))
\label{dhn-s}
\end{eqnarray}
where the $y$ parameter satisfies $0\leq y\leq 1$.
DHN baryons have a valence bound state which can be filled with up to $N$ fermions. In the large $N$ limit, the filling fraction
$\nu=n/N$ becomes a continuous parameter, so that there is in fact a whole one-parameter family of baryons. For self-consistency in the gap equation,  the filling fraction $\nu$ is
related to $y$ by $y=\sin(\pi \nu/2)$. The DHN baryons span the region from
a weakly bound, non-relativistic state at small filling to the ultrarelativistic limit of a decoupled kink and antikink at complete filling. 
At large filling fraction, the DHN baryon looks like a bound kink-antikink molecule, and it is stable not because of topology but because of a balance between the kink-antikink
interaction and the effect of the fermions bound to the kink and antikink. This leads to a direct relation between the baryon size (the distance between the kink and antikink) and the fermion 
filling fraction,  giving a beautiful example of dynamical stability as well as of the Jackiw-Rebbi mechanism of fermion modes bound to localized defects such as kinks \cite{JR}.
This "anatomy" of the DHN baryon is sketched in Figure \ref{f1}.
A particularly interesting special case is the baryon with non-trivial topology, the kink, which is attributed to Callan, Coleman, Gross and Zee (CCGZ) (cited in \cite{L5}). 
Later, Feinberg established the complete set of static solutions to the large $N$ gap equation \cite{L6}, combining inverse scattering theory with resolvent techniques. 
The general solution consists of marginally bound multibaryon configurations whose energy does not depend on the distance between the
constituents. A common feature of all static solutions is the fact that the self-consistent scalar potentials are reflectionless,
generalizing the Kay-Moses potentials of the Schr\"odinger equation \cite{L7} to the Dirac equation \cite{L8}. 

The baryons of the GN model also play a key role in the properties of matter at finite density and temperature. The preferred state of cold matter in the GN model was shown to 
be a crystal, consisting of an array of kinks and antikinks \cite{L9}. The fact that the single baryon potential is reflectionless translates (at finite density)
into a ``finite-gap'' periodic potential, expressed analytically in terms of elliptic functions. The phase diagram as a function of temperature
and chemical potential displays a soliton crystal phase, a massive and a massless Fermi gas phase, and has turned out to be much richer than originally thought \cite{L10,L11}.

Semiclassical methods are not restricted to static solutions. 
Since we are dealing with a relativistic field theory, we can boost any static
solution to an arbitrary Lorentz frame, turning a static Hartree-Fock (HF) solution into a solution of the time-dependent Hartree-Fock (TDHF) approach,
\begin{equation}
\left(i \gamma_5 \partial_x + \gamma^0 S \right)\psi_{\alpha}  = i \partial_t \psi_{\alpha}\quad , \quad S = - g^2 \sum_{\beta}^{\rm occ}
\bar{\psi}_{\beta}\psi_{\beta}.
\label{a2}
\end{equation}
Indeed, it was anticipated already in Witten's seminal paper on baryons in the $1/N$ expansion in QCD \cite{L27}, that  baryon-baryon scattering might be solved with the help 
of the TDHF approach at large $N$. In this paper we realize this goal explicitly for the GN model.
The TDHF approach has already been exploited to compute structure functions of DHN baryons and CCGZ kinks, further confirming the value of the GN model as a
toy model for QCD \cite{L12}. 
In addition to boosted static solutions,  some truly dynamical solutions of Eqs.~(\ref{a2}) are also known. They are harder to find than static HF solutions since
inverse scattering theory is not developed enough for TDHF. The efforts to find non-trivial, time dependent mean field
solutions were also pioneered by DHN who already presented a breather solution, a periodically oscillating (in time) multifermion bound state \cite{L4}.
DHN also pointed out the possibility to relate the breather to the kink-antikink scattering problem by analytic continuation. This suggestion
was taken up and elaborated in several recent works. Kink-antikink scattering was addressed in \cite{L13}.
Apart from a first glimpse of the scattering problem of composite, relativistic objects, the solution also gave several new insights
into the mathematical structure of the theory. A special feature of kink dynamics is the fact that the (valence) fermions do not react
back on the solitons that are carrying them. This decoupling made it possible to formulate kink dynamics in the language of a well-studied, classical
soliton theory, the sinh-Gordon model \cite{L14}, and to use the known $n$-soliton solution of this model to generalize kink-antikink scattering 
to the case of an arbitrary number of kink-like baryons \cite{L15}. From the point of view of TDHF theory, the most striking feature of  
all kink-antikink solutions is the fact that the scalar density of each single-fermion level is proportional to the full self-consistent potential $S$.
This kind of solution where self-consistency holds mode-by-mode was called type I in \cite{L12}. 
This feature in turn opens the way to a geometrical interpretation of HF solutions of the GN model in terms of embedding 2d surfaces into
3d spaces \cite{L16}. Self-consistency translates into the condition of constant mean curvature. One can then understand the relationship 
to minimal surfaces in AdS$_3$ and hence string worldsheets in simple terms, a relation which had been found via the sinh-Gordon equation \cite{L16a}
but was otherwise completely mysterious in \cite{L12}.

As nice as kinks are mathematically, they form only one extreme endpoint of the DHN baryon family. To complete the picture, we need to understand the 
scattering  of general DHN baryons, without the restriction to kinks and antikinks. 
The solution to the scattering problem for two arbitrary DHN baryons would allow one to probe
the degree to which the internal bound state structure is relativistic, all the way
from the non-relativistic limit to the ultrarelativistic one. On the other hand, by choosing the velocity of the baryons, one can cover the
range from non-relativistic to relativistic scattering in the external kinematics as well. This situation is sketched in Figure 2.
\begin{figure}[htb]
\includegraphics[scale=1]{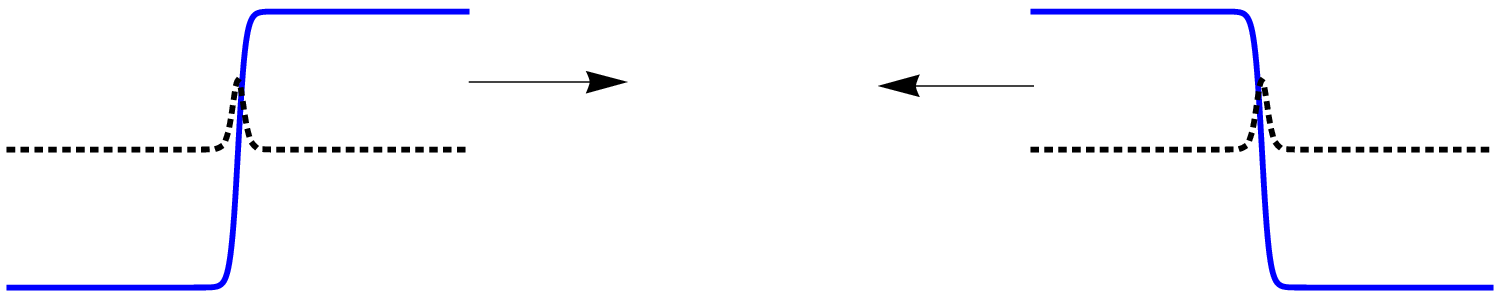}\\
\includegraphics[scale=1]{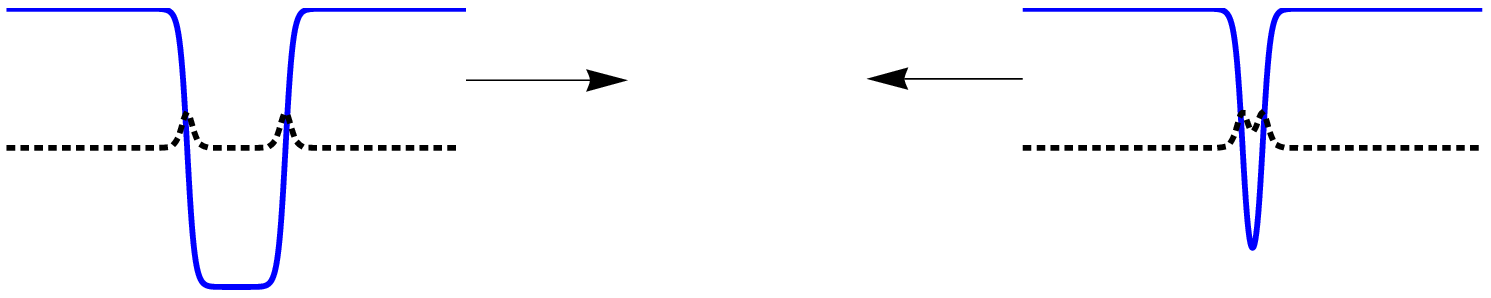}
\caption{The upper figure shows the scattering of a CCGZ-kink baryon and a CCGZ-antikink baryon, as was already discussed in \cite{L13,L15}. In contrast, the lower figure shows the 
problem treated in this current paper: the scattering of two DHN baryons. The solid [blue] lines show the scalar potential, while the dotted [black] lines show the density of the bound 
valence fermions. Note that each asymptotic DHN baryon can have different internal structure (i.e.,  a different $y$ parameter), which affects the baryon size and shape  as well as the 
internal fermion densities; this will be important for understanding the scattering processes. By varying the structural parameters, and also the relative velocity parameter, we can probe 
various interesting physical regimes, spanning continuously from relativistic to non-relativistic.}
\label{f2}
\end{figure}
This general problem is not easy, as evidenced by the fact that it has never been addressed in the literature so far, to the best of our knowledge.
No established, systematic method for solving the TDHF equations is available. Since the valence fermions are now expected to react back,
it is also unlikely that one can map the problem onto some known soliton theory, like sinh-Gordon theory for the kinks. We propose
to solve this problem here by a method based upon an ansatz. As a matter of fact, due to the self-consistency issue, we know neither the equation
nor its solution to start with, so that guessing the right ansatz is quite a challenge. We will explain how a simultaneous ansatz
for the scalar TDHF potential and the (bound and continuum) spinors can nevertheless be found heuristically. The solution of the Dirac equation and the
requirement of self-consistency are sufficient to determine the  unknown parameters of the ansatz and to establish an exact baryon-baryon 
scattering solution in the large $N$ limit of the GN model. The fact that this procedure is successful is undoubtedly related to the
integrability of the GN model at finite $N$ \cite{L17,L18,L19}, although we are not aware of any direct path from the integrability 
of the finite $N$ GN model to the large $N$ solution which we 
find here. Perhaps more important than the specific solution found 
here is the fact that the ansatz can be generalized in a natural way to a whole class of more complicated scattering problems, also
involving multibaryon bound states and breathers in addition to DHN baryons. 

This paper is organized as follows. In Sec.~\ref{sect2}, we illustrate our method of solving the TDHF equations via ansatz in a simpler context,
the single DHN baryon in flight. The results are needed later on, since they enter the scattering problem as incoming and outgoing states.
We also introduce some technicalities related to lightcone variables. Sec.~\ref{sect3} describes the solution of the scattering problem.
We first exhibit the ansatz for the scalar potential, Sec.~\ref{sect3a}, continuum spinors, Sec.~\ref{sect3b}, and bound state spinors, Sec.~\ref{sect3c},
constraining the parameters by asymptotic information whenever possible. Sec.~\ref{sect3d} presents the results for the non-trivial parameters
determined by solving the Dirac equation. In Sec.~\ref{sect3e}, we demonstrate self-consistency of the solution and compute the fermion density.
Sect.~\ref{sect3f} generalizes the results obtained in the center-of-velocity frame to an arbitrary Lorentz frame. In Sec.~\ref{sect4} we report on 
several tests of our results carried out in limiting cases where some information is already available from other sources, i.e., the static limit,
Sec.~\ref{sect4a}, the non-relativistic limit, Sec.~\ref{sect4b}, and the kink-antikink limit, Sec.~\ref{sect4c}. This is followed by several numerical
examples and the relationship to polarons and excitons in conducting polymers in Sec.~\ref{sect4d}. Sec.~\ref{sect5} contains a brief summary and
an outlook. An appendix summarizes the self-consistent scalar potential in conventional variables, complementary to the lightcone formulation used
in the main text.

\section{Single baryon in flight}\label{sect2}

The DHN baryon \cite{L4} is a well-understood multifermion bound state of the large $N$ GN model. Originally derived in the rest frame by
the inverse scattering method, it can easily be boosted to any other inertial frame. We refer to Ref.~\cite{L12} for a detailed discussion of the 
baryon in flight within the Dirac-TDHF approach. In this section, we introduce our ansatz method by pretending that we don't know the 
DHN solution. This illustrates how one can find a self-consistent solution of the Dirac-TDHF equation (\ref{a2}) with a judiciously chosen ansatz for
the scalar potential and the spinors. Apart from this ansatz, 
a crucial ingredient of our approach is the systematic use of lightcone coordinates and variables. These greatly simplify the final results, due
to their simple Lorentz transformation properties. We introduce all the necessary definitions here in a familiar context. The results for the
single baryon in a general Lorentz frame and in the most convenient language are obviously a prerequisite for addressing the two-baryon scattering 
problem, where they appear as incoming and outgoing asymptotic states. 

\subsection{Solution of the Dirac-TDHF equation by ansatz}\label{sect2a}

The key quantity for any Dirac-TDHF calculation in the GN model is the scalar potential $S$. Guided by the idea that the baryon potential has
a kink-antikink shape as well as by covariance, we use as ansatz a rational function of an exponential $U$,
\begin{equation}
S= \frac{\cal N}{\cal D} =  \frac{a_0 + a_1 U+a_{11} U^2}{1+ b_1 U + b_{11} U^2},
\label{1.1}
\end{equation}
with
\begin{equation}
U=\exp \left\{2y\gamma(x-vt)\right\}, \quad \gamma=\frac{1}{\sqrt{1-v^2}}.
\label{1.2}
\end{equation}
$y$ is a parameter governing the size and shape of the potential, $v$ the baryon velocity. Note that for a static ($v=0$) DHN baryon, the solution in (\ref{dhn-s}) can be 
re-written in this form as
\begin{eqnarray}
S(x)=\frac{1+\frac{2(1-2y^2)}{\sqrt{1-y^2}}U+U^2}{1+\frac{2}{\sqrt{1-y^2}}U+U^2}\quad, \quad U=e^{2y\, x}
\label{dhn-s2}
\end{eqnarray}
and it is straightforward to boost this single baryon solution.
Asymptotically, $S$ must reach the value of 
the dynamical fermion mass, set to 1 by our choice of units,
\begin{equation}
\lim_{U \to 0} S = 1, \quad \lim_{U\to \infty} S = 1.
\label{1.3}
\end{equation}
This yields the conditions
\begin{equation}
a_0=1, \quad a_{11}=b_{11}.
\label{1.4}
\end{equation}
By shifting the origin of the $x$ or $t$ axis we can rescale $U$ and hence impose one further condition which we choose as
\begin{equation}
a_{11}=1,
\label{1.5}
\end{equation}
leaving us with two unknown, real coefficients $a_1,b_1$,
\begin{equation}
S= \frac{\cal N}{\cal D} = \frac{1 + a_1 U+ U^2}{1+ b_1 U +  U^2}.
\label{1.6}
\end{equation}
We now turn to the Dirac-TDHF equation (\ref{a2}). We use a chiral representation of the Dirac matrices, $\gamma^0=\sigma_1, \gamma^1 =
 i \sigma_2, \gamma_5 = - \sigma_3$, and introduce lightcone coordinates
\begin{equation}
z=x-t, \quad \bar{z}=x+t, \quad \partial_0 = \bar{\partial}-\partial, \quad \partial_1 = \bar{\partial} + \partial.
\label{1.8}
\end{equation}
$U$ then becomes
\begin{equation}
U = \exp \left\{ y( \eta^{-1} \bar{z} + \eta  z)\right\} ,
\label{1.9}
\end{equation}
where 
\begin{equation}
\eta = e^{\xi} = \sqrt{\frac{1+v}{1-v}}
\label{1.10}
\end{equation}
turns out to be a more convenient variable than either the rapidity $\xi$ or the velocity $v$. The Dirac equation assumes the simple form
\begin{equation}
2i \bar{\partial}  \psi_2 = S \psi_1, \quad 2i \partial \psi_1 = - S \psi_2
\label{1.11}
\end{equation}
in terms of the chiral spinor components $\psi_1= \psi_L, \psi_2= \psi_R$.

In the next step, we try to solve Eqs.~(\ref{1.11}) by an ansatz for the spinors. We assume from the outset that the potential is reflectionless
and that the spinor has the same structure as $S$, multiplied by an exponential (plane wave) factor. In order to have a chance of solving the
Dirac equation, the denominator will be taken to be identical to the one of $S$, whereas we admit arbitrary complex coefficients in 
the numerator. 

Consider the continuum spinors first. They are written as
\begin{equation}
\psi_k  = \left( \begin{array}{c} {\cal N}_1 \\ {\cal N}_2 \end{array} \right)   \frac{e^{i(kx-\omega t)}}{\cal D}, \qquad \omega = \pm \sqrt{1+k^2},
\label{1.12}
\end{equation}
with 
\begin{eqnarray}
{\cal N}_1 & = & c_0 + c_1 U + c_{11} U^2,
\nonumber \\
{\cal N}_2 & = & d_0 + d_1 U + d_{11} U^2,
\label{1.13}
\end{eqnarray}
and ${\cal D}$ from (\ref{1.6}). The exponent can be rephrased in lightcone variables as
\begin{equation}
kx-\omega t = \frac{1}{2}\left( \zeta \bar{z} -\frac{z}{\zeta}\right) 
\label{1.14}
\end{equation}
with the spectral parameter $\zeta$,
\begin{equation}
k= \frac{1}{2} \left( \zeta-\frac{1}{\zeta}\right) , \quad \omega = - \frac{1}{2} \left( \zeta + \frac{1}{\zeta}\right)
\label{1.15}
\end{equation}
(note the more symmetric definition of $\zeta$ as compared to Ref.~\cite{L15}). Like $z$ and $\bar{z}$, $\zeta$ gets simply rescaled under Lorentz
transformations. The sign of $\omega$ is encoded in the sign of $\zeta$, so that all results for continuum states hold for both signs of the
energy. Notice also that it is not necessary to introduce two kinematical variables (lightcone momentum and lightcone energy), since
only on-shell variables enter here. The asymptotics again puts constraints on the parameters of our ansatz for the spinors. Comparison
with the solution of the free, massive ($m=1$) Dirac equation,   
\begin{equation}
\psi^{(0)}_{\zeta}  =  \frac{1}{\sqrt{\zeta^2+1}} \left( \begin{array}{c} \zeta  \\ -1 \end{array} \right) e^{i(\zeta \bar{z} -z/\zeta)/2},
\label{1.16}
\end{equation}
yields 
\begin{eqnarray}
c_0 & = &    \frac{\zeta}{\sqrt{\zeta^2+1}}, \quad c_{11}=T c_0,
\nonumber \\
d_0 & = &  - \frac{1}{\sqrt{\zeta^2+1}}, \quad d_{11}=T d_0.
\label{1.17}
\end{eqnarray}
Here, $c_0$ and $d_0$ are determined by the incident wave, $c_{11}$ and $d_{11}$ by the transmitted wave, $T$ is the unitary transmission
amplitude, $|T|=1$. The complex parameters $c_1,d_1,T$ have to be determined from the Dirac equation, together with the real parameters
$a_1,b_1$ in $S$.

We now insert the ansatz for $S$ and $\psi_{\zeta}$ into the Dirac equation. Since differentiation and multiplication yields again rational 
functions of the exponential $U$, we can equate powers of $U$, reducing the problem to an algebraic one. Owing to the use of lightcone 
variables,  the calculation is straightforward, yielding the following results
\begin{eqnarray}
a_1& = & (1-2 y^2) b_1 , \quad b_1=\frac{2}{w}, \quad w=\sqrt{1-y^2},
\nonumber \\
c_1 & = & \frac{ 2 ( Z^2 - 1 + 2 y^2)}{ w ( Z^2- 2iy Z -1)} c_0, \qquad Z = \eta \zeta,
\nonumber \\
d_1 & = & \frac{- 2 ( 2 Z^2 y^2- Z^2 +1)}{w( Z^2- 2 iy Z -1)} d_0,
\nonumber \\
T & = & \frac{ Z^2+2iy Z -1}{ Z^2 -2iy Z-1}.
\label{1.18}
\end{eqnarray}
$Z$ denotes the spectral parameter $\zeta$, boosted into the rest frame of the baryon. 

Let us consider the bound state spinors next. The transmission amplitude $T$ has 2 poles in the complex $Z$-plane, 
\begin{equation}
Z^2-2iy Z -1=(Z - Z_1)(Z + Z_1^*), \quad Z_1 = iy - w
\label{1.19}
\end{equation}
corresponding to the positive and negative energy bound states at $E= \pm w$ in the rest frame of the baryon.
Since the potential is transparent, the spinor may be regarded as a continuum spinor, analytically continued
to complex $\zeta$. For positive energy spinors, this suggests the ansatz
\begin{equation}
\psi^{(1)}  = \left( \begin{array}{c} {\cal N}_1^{(1)} \\ {\cal N}_2^{(1)} \end{array} \right)   \frac{e^{ i {\cal F}}}{\cal D}
\label{1.20}
\end{equation}
with the exponent
\begin{equation}
{\cal F} = iy \gamma (x-vt) - w \gamma (t-vx) = \frac{1}{2} \left( \zeta_1\bar{z} -  \frac{z}{\zeta_1}\right), \quad \zeta_1= \eta^{-1} Z_1
\label{1.21}
\end{equation}
and
\begin{eqnarray}
{\cal N}_1^{(1)} & = & e_1 U + e_{11} U^2,
\nonumber \\
{\cal N}_2^{(1)} & = &  f_1 U + f_{11} U^2.
\label{1.22}
\end{eqnarray}
Since 
\begin{equation}
|e^{i{\cal F}}|^2 = U^{-1},
\label{1.22a}
\end{equation}
constant terms with $e_0, f_0$ have to be omitted in the numerators (\ref{1.22}) for the sake of normalizability. If one inserts this ansatz
into the Dirac equation, one can determine the unknown parameters algebraically except for the overall normalization with the result
\begin{eqnarray}
e_{11} & = & - Z_1 e_1,
\nonumber \\
f_1 & = & -  Z_1 \eta e_1,
\nonumber \\
f_{11} & = & \eta e_1.
\label{1.23}
\end{eqnarray}
The parameter $e_1$ then follows from the normalization condition which we choose in the non-covariant form
\begin{equation}
\int dx (\psi^{\dagger} \psi)^{(1)} = 1,
\label{1.23a}
\end{equation}
yielding
\begin{equation}
e_1^2 = \frac{y}{\eta w}  .
\label{1.24}
\end{equation}
Before proceeding, we have to introduce one further element of lightcone notation. One advantage of using $\zeta$ and $\eta$ rather
than $k$ and $v$ is the fact that all square roots of the type $\sqrt{1+k^2},\sqrt{1-v^2}$ are eliminated. In order to avoid also $w=\sqrt{1-y^2}$,
it is advantageous to combine $y,w$ into the complex variable $Z_1$ introduced in (\ref{1.19}). Using the fact that $|Z_1|=1$, we get
\begin{equation}
y = \frac{Z_1^2-1}{2i Z_1}, \quad w = - \frac{Z_1^2+1}{2 Z_1}.
\label{1.25}
\end{equation}
In this new notation the relevant single baryon quantities read
\begin{eqnarray}
a_1 & = & - \frac{2(Z_1^4+1)}{Z_1(Z_1^2+1)},
\nonumber \\
b_1 & = & - \frac{4 Z_1}{Z_1^2+1},
\nonumber \\
c_1 & = &  - \frac{2(2 Z^2 Z_1^2 - Z_1^4-1)}{(Z_1^2+1)(Z  - Z_1)( Z Z_1+1)} c_0,
\nonumber \\
d_1 & = &  - \frac{ 2 ( Z^2  Z_1^4 +  Z^2 - 2 Z_1^2)}{(Z_1^2+1)(Z -Z_1)(Z Z_1+1)} d_0,
\nonumber \\
T & = & \frac{(Z + Z_1)( Z Z_1 -1)}{(Z - Z_1)(Z Z_1 +1)},
\nonumber \\
e_1^2 & = &  \frac{i}{\eta} \left( \frac{Z_1^2-1}{Z_1^2+1} \right) .
\label{1.26}
\end{eqnarray}
All other coefficients are the same as in Eqs.~(\ref{1.17}) and (\ref{1.23}). A minor shortcoming of this notation is the fact that the unitarity of
$Z_1$ ($|Z_1|=1$) is not manifest. Thus for instance, the transmission amplitude $T$ could also be represented as
\begin{equation}
T= \frac{(Z-Z_1^*)(Z+Z_1)}{(Z-Z_1)(Z+Z_1^*)}
\label{1.27}
\end{equation}
since $1/Z_1=Z_1^*$. This would show at once the poles of the positive ($Z=Z_1$) and negative ($Z=-Z_1^*$) energy bound states, as well
as the fact that $|T|=1$. To ease the notation of more complicated expressions later on however, we prefer not to use complex conjugate
variables and will stick to the notation of Eqs.~(\ref{1.26}).

Notice that one could also generate the bound state spinor by analytic continuation from the continuum spinors 
(up to the normalization factor), rather than by ansatz. As one can easily check, it is sufficient to evaluate the residue of $\psi_{\zeta}$ at the pole $\zeta=\zeta_1$. 
Finally, the negative energy bound state spinor can trivially be obtained from the positive energy one by the mapping $\psi_1 \to \psi_1^*, 
\psi_2 \to - \psi_2^*$, so that there is no need to discuss it separately.

\subsection{Self-consistency and fermion density}\label{sect2b}

Having determined the free parameters in the ansatz for $S$ and solved the Dirac equation, we proceed to 
verify self-consistency and derive the relation between the parameter $y$ and the fermion number of the DHN baryon. To this end, we first
need the scalar density $\bar{\psi}\psi$ for all occupied states. We decompose the scalar density for continuum states computed from the
above spinors as follows,
\begin{eqnarray}
\bar{\psi}_{\zeta}\psi_{\zeta}  &  = &  (\bar{\psi}_{\zeta}\psi_{\zeta})_1   + (\bar{\psi}_{\zeta}\psi_{\zeta})_2,
\nonumber \\
(\bar{\psi}_{\zeta}\psi_{\zeta})_1 & = & - \frac{2\zeta}{\zeta^2+1} S ,
\nonumber \\
(\bar{\psi}_{\zeta}\psi_{\zeta})_2   & = &  \frac{2 \zeta Z^2 (Z_1^2+1)^2}{(\zeta^2+1) (Z^2 Z_1^2-1)( Z^2-Z_1^2)}(1-S) .
\label{1.28}
\end{eqnarray}
Positive and negative energy bound states on the other hand yield
\begin{equation}
\bar{\psi}\psi  =   \mp \frac{i}{2} \left( \frac{Z_1^2+1}{Z_1^2-1}\right) (1-S).
\label{1.29}
\end{equation}
The scalar density of any occupied state can be written as a linear combination of two independent functions of ($x,t$), so that we are dealing
with a type II solution of the TDHF problem. In the case of the kink, $y=1, Z_1=i$ so that ($\bar{\psi}_{\zeta} \psi_{\zeta})_2$ and the bound state
scalar density vanish and the solution becomes type I.

The self-consistency relation in the massless Gross-Neveu model is
\begin{equation}
\langle \bar{\psi}\psi \rangle  =  - \frac{1}{Ng^2} S.
\label{1.30}
\end{equation}
Renormalization of the coupling constant is performed with the help of the vacuum gap equation ($S=1$), using a cutoff regularization
\begin{equation}
\int_{- \Lambda/2}^{\Lambda/2} \frac{dk}{2\pi} \to \int_{1/\Lambda}^{\Lambda} \frac{d\zeta}{2\pi} \frac{\zeta^2+1}{2\zeta^2}, 
\quad (\bar{\psi}\psi)_{\zeta}^{(0)}= -  \frac{2\zeta}{\zeta^2+1}.
\label{1.31}
\end{equation}
We find
\begin{eqnarray}
\frac{1}{Ng^2} & = &  \frac{1}{\pi} \ln \Lambda 
\label{1.32}
\end{eqnarray}
so that Eq.~(\ref{1.30}) becomes
\begin{equation}
\langle \bar{\psi} \psi \rangle = - \frac{S}{\pi} \ln \Lambda .
\label{1.33}
\end{equation}
The first contribution to the chiral condensate from the continuum in Eq.~(\ref{1.28}) already gives self-consistency,
\begin{equation}
\int_{1/\Lambda}^{\Lambda} \frac{d\zeta}{2\pi} \frac{\zeta^2+1}{2\zeta^2} (\bar{\psi}_{\zeta} \psi_{\zeta})_1 = - \frac{S}{\pi} \ln \Lambda.
\label{1.34}
\end{equation}
Hence the 2nd contribution, 
\begin{equation}
\int_{0}^{\infty} \frac{d\zeta}{2\pi} \frac{\zeta^2+1}{2\zeta^2}  (\bar{\psi}_{\zeta} \psi_{\zeta})_2 = \frac{S-1}{4\pi} \left( \frac{Z_1^2+1}
{Z_1^2-1}\right) \ln Z_1^4,
\label{1.35}
\end{equation}
must cancel the contributions from the discrete states (occupation fractions $\nu_{\pm}$),   
\begin{equation}
2 \pi i (\nu_+ - \nu_-)   +  \ln Z_1^4 = 0 .
\label{1.36}
\end{equation}
This determines the parameter $y$, given the occupation of the bound states.
For the ground state baryons ($\nu_-=1$) in particular, we recover the 
result of DHN,
\begin{equation}
y = \sin \left( \frac{\pi \nu_+}{2} \right).
\label{1.38}
\end{equation}
In order to relate the occupation fractions $\nu_{\pm}$ of the bound states to the fermion number of the baryon, we still have to compute the
fermion density, including the induced contribution from the Dirac sea. The density of the continuum states can be expressed through $S$ 
as follows,
\begin{equation}
\psi_{\zeta}^{\dagger} \psi_{\zeta} = 1 - \frac{(Z^2+1)(\zeta^2+ Z^2)Z_1^2}{(\zeta^2+1)(Z^2 Z_1^2 -1)(Z^2-Z_1^2)} (1-S^2).
\label{1.40}
\end{equation}
Subtracting the divergent density of the non-interacting Dirac sea, this yields the following contribution to the fermion density, 
\begin{equation}
\int_0^{\infty}  \frac{d\zeta}{2\pi} \frac{\zeta^2+1}{2\zeta^2}   (\psi_{\zeta}^{\dagger} \psi_{\zeta}   -1)  =  - \frac{i Z_1 (\eta^2 + 1)}{4\eta (Z_1^2-1)}(1-S^2).
\label{1.41}
\end{equation}
Positive and negative energy bound states give the density
\begin{equation}
(\psi^{\dagger}\psi)^{(1)}  =   \frac{i Z_1 (\eta^2 + 1)}{4\eta (Z_1^2+1)}(1-S^2).  
\label{1.42}
\end{equation}
The total (subtracted) fermion density of the DHN baryon can therefore be expressed in terms of the bound state density as
\begin{equation}
\langle \psi^{\dagger}\psi \rangle   =  \left( \nu_+ + \nu_- - 1 \right) (\psi^{\dagger}\psi)^{(1)},
\label{1.43}
\end{equation}
whereas the fermion number becomes 
\begin{equation}
N_f  =  N(\nu_+ + \nu_- -  1).
\label{1.43a}
\end{equation}
If $\nu_-=1$ (ground state baryon), the induced fermion density from the continuum and the fermion density from the negative 
energy bound state cancel exactly, so that one can read off the full fermion density from the positive valence level alone. 

We finish this section with a comment on the massive GN model, i.e., Lagrangian (\ref{a1}) supplemented by a mass term $- m_0 \bar{\psi}\psi$.
It is well known that the functional form of $S$ of the DHN baryon also solves the massive GN model, the only difference being the relationship 
between $y$ and $\nu_{\pm}$ \cite{L20,L21}. The self-consistency condition (\ref{1.30}) has to be replaced by
\begin{equation}
\langle \bar{\psi}\psi \rangle  =  - \frac{1}{Ng^2} (S-m_0) = - \frac{1}{Ng^2} S  + \frac{\gamma_c}{\pi}
\label{1.44}
\end{equation}
with the ``confinement parameter"
\begin{equation}
\gamma_c = \frac{\pi m_0}{Ng^2}.
\label{1.45}
\end{equation}
The vacuum gap equation (\ref{1.32}) now reads
\begin{equation}
\frac{1}{Ng^2}  =   \frac{1}{\pi} (\gamma_c + \ln \Lambda )
\label{1.46}
\end{equation} 
so that (\ref{1.44}) becomes
\begin{equation}
\langle \bar{\psi}\psi \rangle  = - \frac{S}{\pi} \ln \Lambda  + \frac{\gamma_c
}{\pi}(1-S)
\label{1.47}
\end{equation}
It can be satisfied in the case of the DHN baryon because $\langle \bar{\psi}\psi \rangle + (S/\pi)\ln \Lambda$ (which receives contributions
from the continuum and the bound states) is proportional to $(1-S)$.

\section{Baryon-baryon scattering}\label{sect3}

We now apply the ansatz method to a more difficult problem where the solution is not known --- scattering of two DHN baryons with
different baryon numbers (parameters $y_1,y_2$). To keep the number of parameters as small as possible, we shall work in the 
center-of-velocity frame where the baryon velocities are $\pm v$. Since the calculation  is fully covariant, we can transform the results into
any other Lorentz frame afterwards. In analogy to the one baryon problem, we parameterize the scalar potential as a rational function 
of exponentials. The spinors are again taken to be exponentials times functions similar to $S$, where we always insist on keeping the same 
denominators. In the single baryon case, the asymptotic information which we used to reduce the number of parameters came from the 
vacuum. Now we will similarly exploit the asymptotic information from the incoming and outgoing baryons. This recursive way of proceeding
greatly reduces the number of parameters which then have to be determined algebraically via the Dirac equation. 

\subsection{Ansatz for scalar potential}\label{sect3a}

Our ansatz for the scalar potential is 
\begin{equation}
S = \frac{\cal N}{\cal D}
\label{2.1}
\end{equation}
with
\begin{equation}
{\cal N}  =  1 + a_1 U_1 + a_2 U_2 + a_{11} U_1^2 + a_{12} U_1 U_2 + a_{22} U_2^2 + a_{112} U_1^2 U_2 + a_{122} U_1 U_2^2
+ a_{1122} U_1^2 U_2^2
\label{2.2}
\end{equation}
\begin{equation}
{\cal D}  =  1 + b_1 U_1 + b_2 U_2 + b_{11} U_1^2 + b_{12} U_1 U_2 + b_{22} U_2^2 + b_{112} U_1^2 U_2 + b_{122} U_1 U_2^2
+ b_{1122} U_1^2 U_2^2
\label{2.3}
\end{equation}
and
\begin{equation}
U_1 = \exp \left\{2y_1 \gamma (x-vt)\right\}, \qquad U_2 = \exp \left\{2y_2\gamma (x+vt)\right\}
\label{2.4}
\end{equation}
The $U_1,U_2$ dependence is motivated by the product of the two baryon potentials, which we must recover when the 
scatterers are well separated. In this sense, the ansatz is the minimal one having a chance of describing baryon-baryon scattering.
Almost all of the 16 real parameters in $S$ are determined by the asymptotic in- and out-states as follows. For $t\to -\infty$, $S$ reduces
to 
\begin{equation}
\lim_{U_2 \to 0}  S = \frac{1+a_1 U_1 + a_{11} U_1^2}{1+b_1 U_1 + b_{11}U_1^2}
\label{2.5}
\end{equation}
near incoming baryon 1 and to
\begin{equation}
\lim_{U_1 \to \infty} S =  \frac{a_{11}+a_{112} U_2 + a_{1122} U_2^2}{b_{11}+b_{112} U_2 + b_{1122}U_2^2}
\label{2.6}
\end{equation}
near incoming baryon 2. For $t\to \infty$, we get the limit
\begin{equation}
\lim_{U_2 \to \infty} S = \frac{a_{22}+a_{122} U_1 + a_{1122} U_1^2}{b_{22}+b_{122} U_1 + b_{1122}U_1^2}
\label{2.7}
\end{equation}
near outgoing baryon 1 and 
\begin{equation}
\lim_{U_1 \to 0}S =  \frac{1+a_{2} U_2 + a_{22} U_2^2}{1+b_{2} U_2 + b_{22}U_2^2}
\label{2.8}
\end{equation}
near outgoing baryon 2. These expressions should be matched to the single baryon scalar potential (\ref{1.6}). In this step, we have to account
for the time delay occuring during the collision. To this end, we replace $U_1\to U_1/\delta_{12}, U_2 \to U_2/\delta_{21}$ (with real 
$\delta_{12},\delta_{21}$) in the single baryon scalar potentials of the outgoing baryons. We only get a consistent parameterization for 
$\delta_{12} = 1/\delta_{21} := \delta$ and can now fix all coefficients in $S$ except for $a_{12},b_{12}$ 
and $\delta$ as follows, \begin{eqnarray}
a_1 & = & a_1^I, \qquad a_2 = a_1^{II} \delta, 
\nonumber \\
a_{112} & = & a_1^{II}, \qquad a_{122} = a_1^I \delta,
\nonumber \\
b_1 & = & b_1^I, \qquad b_2= b_1^{II} \delta,
\nonumber \\
b_{112} & = & b_1^{II}, \qquad b_{122}  =  b_1^I \delta,
\nonumber \\
a_{11} & = & a_{1122} = b_{11} = b_{1122} = 1,
\nonumber \\
a_{22} & = & b_{22}  = \delta^2.
\label{2.9}
\end{eqnarray}
The superscripts $I,II$ on the one-baryon coefficients $a_1,b_1$ on the right side of these equations refer to baryons $I$ (parameters $y_1,v$)
and $II$ (parameters $y_2,-v$), respectively. As in the one-baryon case, it is advisable to use lightcone variables to keep the results for the 
spinors in a manageable form. We therefore write $U_{1,2}$ as
\begin{eqnarray}
U_1 & = & \exp \left\{ y_1(\eta^{-1} \bar{z} + \eta z )\right\},
\nonumber \\
U_2 & = &  \exp \left\{ y_2(\eta \bar{z} + \eta^{-1} z )\right\},
\label{2.13}
\end{eqnarray}
with $\eta$ as defined in (\ref{1.10}), and use the parameterization ($i=1,2$)
\begin{eqnarray}
Z_i & =&    i y_i-w_i, \qquad |Z_i|=1,
\nonumber \\
y_i & = &  \frac{Z_i^2-1}{2iZ_i}, \qquad w_i  =  - \frac{Z_i^2+1}{2 Z_i}.
\label{2.11}
\end{eqnarray}
The single baryon coefficients entering Eqs.~(\ref{2.9}) then go over into 
\begin{eqnarray}
a_1^I & = & - \frac{2(Z_1^4+1)}{Z_1(Z_1^2+1)}, \qquad b_1^I = -\frac{4 Z_1}{Z_1^2+1},
\nonumber \\
a_1^{II} & = & - \frac{2(Z_2^4+1)}{Z_2(Z_2^2+1)}, \qquad b_1^{II} = -\frac{4 Z_2}{Z_2^2+1}.
\label{2.12}
\end{eqnarray}
We will see later that $\delta$ can actually be predicted on the basis of the single baryon input. This leaves us with only two unknown, real parameters
$a_{12}, b_{12}$, to be determined from the Dirac-TDHF equation.

\subsection{Ansatz for continuum spinors}\label{sect3b}

Anticipating that the mean field is reflectionless, we propose the following ansatz for the continuum spinor in TDHF,
\begin{equation}
\psi_{\zeta}  = \left( \begin{array}{c} {\cal N}_1 \\ {\cal N}_2 \end{array} \right)   \frac{e^{i(\zeta \bar{z}-z/\zeta)/2}}{\cal D}.
\label{2.14}
\end{equation}
$\zeta$ is the spectral parameter introduced in (\ref{1.15}), ${\cal D}$ is the denominator of $S$, and the numerators are polynomials in 
$U_1,U_2$ of the same degree as ${\cal D}$, 
\begin{equation}
{\cal N}_1  =  c_0 + c_1 U_1 + c_2 U_2 + c_{11} U_1^2 + c_{12} U_1 U_2 + c_{22} U_2^2 + c_{112} U_1^2 U_2 + c_{122} U_1 U_2^2
+ c_{1122} U_1^2 U_2^2,
\label{2.15}
\end{equation}
\begin{equation}
{\cal N}_2  =  d_0 + d_1 U_1 + d_2 U_2 + d_{11} U_1^2 + d_{12} U_1 U_2 + d_{22} U_2^2 + d_{112} U_1^2 U_2 + d_{122} U_1 U_2^2
+ d_{1122} U_1^2 U_2^2.
\label{2.16}
\end{equation}
Most of the parameters are again determined by asymptotics. For the incoming baryon $I$ and the outgoing baryon $II$, we can literally 
follow the treatment of $S$ by letting $U_{1,2} \to 0$, respectively, with the result
\begin{eqnarray}
c_0 & = & c_0^I, \quad c_1 =  c_1^I, \quad c_{11} = c_{11}^I,
\nonumber \\
d_0 & = & d_0^I, \quad d_1 = d_1^I, \quad d_{11} = d_{11}^I,
\nonumber \\
c_0 & = & c_0^{II}, \quad c_2 = \delta c_1^{II}, \quad c_{22} = \delta^2 c_{11}^{II},
\nonumber \\
d_0 & = & d_0^{II}, \quad d_2 = \delta d_1^{II}, \quad d_{22} = \delta^2 d_{11}^{II}.
\label{2.17}
\end{eqnarray}
We have taken into account the time delay for baryon $II$ through the substitution $U_2 \to U_2/\delta$. For incoming baryon $II$ and 
outgoing baryon $I$, we must account for the time delay of baryon $I$ ($U_1 \to \delta U_1$). In addition, here the incident continuum states have  
already been scattered by the other baryon. We therefore insert single baryon transmission amplitudes $T^{I,II}$ into the results analogous
to $S$ 
(for $U_{1,2}\to \infty$), 
\begin{eqnarray}
c_{11} & = & T^I c_0^{II}, \quad c_{112} = T^I c_1^{II}, \quad c_{1122} = T^I c_{11}^{II},
\nonumber \\
d_{11} & = & T^I d_0^{II}, \quad d_{112} = T^I d_1^{II}, \quad d_{1122} = T^I d_{11}^{II},
\nonumber \\
c_{22} & = & T^{II} \delta^2 c_0^I, \quad c_{122} = T^{II} \delta c_1^I, \quad c_{1122} = T^{II}  c_{11}^I,
\nonumber \\
d_{22} & = & T^{II} \delta^2 d_0^I, \quad d_{122} = T^{II} \delta d_1^I, \quad d_{1122} = T^{II} d_{11}^I.
\label{2.18}
\end{eqnarray}
For convenience we list all single baryon parameters entering Eqs.~(\ref{2.17},\ref{2.18}). For baryon $I$, take over the results
from Sec.~\ref{sect2a}, with the notation
\begin{equation}
Z \to Z_I = \eta \zeta,  \quad Z_1 = i y_1-w_1,
\label{2.19}
\end{equation}
i.e.,
\begin{eqnarray}
c_0^I & = & \frac{\zeta}{\sqrt{\zeta^2+1}}, \qquad c_{11}^I  =  T^I c_0^I,
\nonumber \\
c_1^I & = & - \frac{2(2 Z_I^2 Z_1^2-Z_1^4-1)}{(Z_1^2+1)(Z_I-Z_1)(Z_IZ_1+1)}c_0^I,
\nonumber \\
d_0^I & = & -\frac{1}{\sqrt{\zeta^2+1}}, \qquad d_{11}^I = T^I d_0^I,
\nonumber \\
d_1^I & = & - \frac{2(Z_I^2 Z_1^4+Z_I^2-2Z_1^2)}{(Z_1^2+1)(Z_I-Z_1)(Z_IZ_1+1)}d_0^I.
\label{2.20}
\end{eqnarray}
The analogous expressions for baryon $II$ are
\begin{equation}
Z \to Z_{II} = \eta^{-1} \zeta,  \quad Z_2 = i y_2-w_2
\label{2.21}
\end{equation}
and 
\begin{eqnarray}
c_0^{II} & = & \frac{\zeta}{\sqrt{\zeta^2+1}}, \qquad c_{11}^{II}  =  T^{II} c_0^{II},
\nonumber \\
c_1^{II} & = & - \frac{2( 2 Z_{II}^2 Z_2^2-Z_2^4-1)}{(Z_2^2+1)(Z_{II}-Z_2)(Z_{II}Z_2+1)}c_0^{II},
\nonumber \\
d_0^{II} & = & -\frac{1}{\sqrt{\zeta^2+1}}, \qquad d_{11}^{II} = T^{II}d_0^{II},
\nonumber \\
d_1^{II} & = & - \frac{2(Z_{II}^2 Z_2^4+Z_{II}^2-2Z_2^2)}{(Z_2^2+1)(Z_{II}-Z_2)(Z_{II} Z_2+1)}d_0^{II}.
\label{2.22}
\end{eqnarray}
The transmission amplitudes can be inferred from Eqs.~(\ref{1.26}),
\begin{eqnarray}
T^I & = &  \frac{(Z_I+ Z_1)(Z_I Z_1 -1)}{(Z_I- Z_1)(Z_I Z_1 +1)}
\nonumber \\
T^{II} & = &   \frac{(Z_{II}+ Z_2)(Z_{II} Z_2 -1)}{(Z_{II}- Z_2)(Z_{II} Z_2 +1)}
\label{2.23}
\end{eqnarray}
Some coefficients appear repeatedly in Eqs.~(\ref{2.17},\ref{2.18}). Due to the single baryon identities
\begin{equation}
c_0^I  =  c_0^{II}, \quad c_{11}^I = T^I c_0^I, \quad c_{11}^{II} = T^{II} c_0^{II}
\label{2.24}
\end{equation}
and similar relations for $d$'s, there is no conflict though. Eventually, only two complex parameters in the continuum spinor remain to be determined,
$c_{12}$ and $d_{12}$. Finally, we note that the transmission amplitude of the full continuum spinor factorizes,
\begin{equation}
T = T^I T^{II}.
\label{2.25}
\end{equation}
This simply follows from the fact that we may evaluate it at a time where the baryons are well separated.

\subsection{Ansatz for bound state spinors}\label{sect3c}

It is sufficient to consider the two positive energy bound states. Along the lines discussed above, we set
\begin{equation}
\psi^{(i)}  = \left( \begin{array}{c} {\cal N}_1^{(i)} \\ {\cal N}_2^{(i)} \end{array} \right)   \frac{e^{i{\cal F}^{(i)}}}{\cal D} , \qquad (i=1,2)
\label{2.26}
\end{equation}
where ${\cal D}$ is the denominator of $S$. 
Let us focus on the first bound state (asymptotically belonging to baryon $I$), since the 2nd one can simply be obtained by a relabeling of
baryons $I$ and $II$. As ansatz for the numerators, we choose,
\begin{eqnarray}
{\cal N}_1^{(1)} &  = &  e_1 U_1  + e_{11} U_1^2 + e_{12} U_1 U_2 + e_{112} U_1^2 U_2 +  e_{122} U_1 U_2^2 + e_{1122} U_1^2 U_2^2,
\nonumber \\
{\cal N}_2^{(1)} &  = & f_1 U_1  + f_{11} U_1^2 + f_{12} U_1 U_2 + f_{112} U_1^2 U_2 +  f_{122} U_1 U_2^2 + f_{1122} U_1^2 U_2^2.
\label{2.27}
\end{eqnarray}
The exponent is determined by the known bound state energy and kinematics,
\begin{equation}
{\cal F}^{(1)}  =   \frac{1}{2} \left( \zeta_1 \bar{z} - \frac{z}{\zeta_1} \right), \quad \zeta_1 = \eta^{-1} Z_1
\label{2.28}
\end{equation}
with $Z_1$ from Eq.~(\ref{2.19}). Note that
\begin{equation}
|e^{i {\cal F}^{(1)}}|^2 = U_1^{-1}.
\label{2.29}
\end{equation}
Normalizability then forces us to leave out all terms not containing $U_1$ in (\ref{2.27}). 

We now turn to the constraints from the asymptotic behavior of $\psi^{(1)}$. Since bound state 1 is attached to baryon $I$ asymptotically, 
the only issues are the incoming and outgoing baryon $I$. For incoming baryon $I$, let $U_2 \to 0$ and find 
\begin{equation}
e_{1}   =    e_{1}^I, \quad e_{11}  =  e_{11}^I, \quad f_{1}  =  f_{1}^I, \quad f_{11}  =  f_{11}^I.
\label{2.30}
\end{equation}
In the outgoing channel, let $U_2\to \infty$. Here the bound state spinor 1 acquires a transmission amplitude from baryon $II$ denoted
as $T_1^{II}$. The asymptotic conditions therefore read
\begin{equation}
e_{122}  =  T^{II}_1 \delta e_1^I, \quad e_{1122}  = T^{II}_1 e_{11}^I, \quad f_{122}  =  T^{II}_1  \delta f_1^I, \quad f_{1122}  = T^{II}_1  f_{11}^I.
\label{2.31}
\end{equation}
Let us collect the required one-baryon coefficients,
\begin{eqnarray}
e_{11}^I & = &  - Z_1 e_1^I,
\nonumber \\
f_1^I & = & - Z_1 \eta e_1^I,
\nonumber \\
f_{11}^I & = & \eta e_1^I,
\nonumber \\
(e_1^I)^2 & = & \frac{i}{\eta} \left( \frac{Z_1^2-1}{Z_1^2+1}\right) .
\label{2.32}
\end{eqnarray}
How does one compute a transmission amplitude for a bound state? Since the bound state spinor can be thought of as a continuum 
spinor with complex spectral parameter, we can find $T_1^{II}$ by analytic continuation. Take $T^{II}$ from  Eq.~(\ref{2.23}) and replace the 
argument $Z_{II}$ by $e^{-2\xi}Z_1 = \eta^{-2} Z_1$. $e^{-2\xi}$ appears because we have to boost from the rest frame of baryon $I$ 
(velocity $v$) to the rest frame of baryon $II$ (velocity $-v$). In this way we arrive at
\begin{equation}
T_1^{II}  =   \frac{( \eta^{-2} Z_1 + Z_2)(\eta^{-2} Z_1 Z_2 -1)}{(\eta^{-2} Z_1-  Z_2)(\eta^{-2} Z_1 Z_2 +1)}.
\label{2.33}
\end{equation}
Since the spectral parameter is now complex, $T_1^{II}$ is not unitary. Its phase produces a phase shift of the bound state spinor, 
whereas its modulus gives rise to a time delay. This last observation gives us a clue for computing the time delay factor $\delta$ introduced
in the ansatz for $S$. Since the bound state spinor is moving along with the baryon asymptotically and the potential is quadratic in the 
spinor, $\delta$ must satisfy the relation
\begin{equation}
\delta = |T_1^{II}|^2.
\label{2.34}
\end{equation}
Otherwise, there would be a spatial shift between the scalar potential and the bound state spinor of the scattered baryon.
It is surprising that one can predict the time delay, a two-baryon scattering observable, on the basis of single baryon results alone. 
Evaluating (\ref{2.34}) yields
\begin{equation}
\delta = \frac{\hat{n}}{\hat{d}} = \frac{( \eta^2 Z_2+ Z_1)( \eta^2  Z_1+ Z_2)( \eta^2  Z_1 Z_2-1)( \eta^2 - Z_1 Z_2)}
{( \eta^2  Z_2- Z_1)( \eta^2  Z_1- Z_2)(\eta^2 Z_1 Z_2+1)( \eta^2 + Z_1 Z_2)}.
\label{2.35}
\end{equation}
Eqs.~(\ref{2.30}--\ref{2.35}) determine the parameters of the bound state spinor except for the 4 complex coefficients 
$e_{12},e_{112},f_{12},f_{112}$. 

\subsection{Nontrivial coefficients from the solution of the Dirac equation}\label{sect3d}

The ansatz for $S$ and the spinors has been constrained as much as possible by single baryon results and asymptotics.
This reduces the number of parameters  from 16 to 2 real parameters in $S$, from 18 to 2 complex
parameters in the continuum spinor, and from 12 to 4 complex parameters in the bound state 
spinor. The remaining coefficients have to be determined by inserting everything into the Dirac equation (\ref{1.11})
and equating coefficients of monomials $U_1^n U_2^m$. The system is strongly overdetermined, so that the existence of a solution 
is non-trivial. We have indeed found a unique solution, confirming the correctness of our ansatz, with the following parameters:
\subsubsection{Scalar potential}\label{sect3d1}
The non-trivial parameters $a_{12}, b_{12}$ are given by
\begin{eqnarray}
a_{12} & = & \frac{ 4 Z_1^2 Z_2^2 (Z_2^4+1)(Z_1^4+1)(\eta^8+1) -  8 (Z_1^4 Z_2^4+1)(Z_1^4+Z_2^4)\eta^4 }{Z_1 Z_2 (Z_1^2+1)(Z_2^2+1)\hat{d}},
\nonumber \\
b_{12} & = & \frac{8 Z_1 Z_2 \left[ 2 Z_1^2 Z_2^2 (\eta^8+1)-  (Z_1^4+1)(Z_2^4+1)\eta^4 \right]}{(Z_1^2+1)(Z_2^2+1)\hat{d}},
\label{2.37}
\end{eqnarray}
with $\hat{d}$ from Eq.~(\ref{2.35}).
\subsubsection{Continuum spinor}\label{sect3d2}
The parameters $c_{12},d_{12}$ are found to be
\begin{equation}
c_{12} =  \frac{4 \zeta \hat{c}_{12}}{\sqrt{1+ \zeta^2} D_{12}}, \quad d_{12} = \frac{\hat{d}_{12}}{\sqrt{1+ \zeta^2} D_{12}},
\label{2.39}
\end{equation} 
with the common denominator 
\begin{eqnarray}
D_{12} & = &  (Z_1^2+1)(Z_2^2+1)(\eta^2 Z_1 Z_2 +1)(Z_1 \eta^2-Z_2)(\eta^2+Z_1Z_2)(Z_1-\eta^2 Z_2)
\nonumber \\
& & (\zeta-\eta Z_2)(\zeta Z_2+\eta)(\eta \zeta-Z_1)(\eta \zeta Z_1+1)
\label{2.40}
\end{eqnarray}
and the numerators
\begin{eqnarray}
\hat{c}_{12} & = & 2 \zeta^2 Z_1^2 Z_2^2 \left[ Z_2^2 (Z_1^4+1) + Z_1^2  (Z_2^4+1) \eta^{12} \right]
\nonumber \\
& & - \eta^2 (1+\eta^8) Z_1^2 Z_2^2 \left[ (1+Z_1^4)(1+Z_2^4)+ 4 Z_1^2 Z_2^2 \zeta^4 \right]
\nonumber \\
& & - \zeta^2 \eta^4 \left[ Z_2^2 (Z_2^4+1)(Z_1^8+1)+ \eta^4 Z_1^2 (Z_1^4+1)(Z_2^8+1) \right]
\nonumber  \\
& & +  2 \eta^6 \left[ (Z_1^4+Z_2^4)(Z_1^4 Z_2^4+1) + \zeta^4 Z_1^2 Z_2^2 (Z_1^4+1)(Z_2^4+1) \right],
\nonumber \\
\hat{d}_{12} & = & - 8 \zeta^2 Z_1^2 Z_2^2 \left[ Z_1^2 (Z_2^4+1) + \eta^{12}  Z_2^2 (Z_1^4+1) \right]
\nonumber \\
& & + 4 Z_1^2 Z_2^2 \eta^2 (1 +  \eta^8) \left[ \zeta^4(Z_1^4+1)(Z_2^4+1)+ 4 Z_1^2 Z_2^2 \right]
\nonumber \\
& & + 4 \zeta^2 \eta^4 \left[ Z_1^2(Z_1^4+1)(Z_2^8+1)+ \eta^4 Z_2^2(Z_1^8+1)(Z_2^4+1) \right]
\nonumber \\
& &  - 8 \eta^6 \left[ Z_1^2 Z_2^2 (Z_1^4+1)(Z_2^4+1) + \zeta^4 (Z_1^4+Z_2^4)(Z_1^4 Z_2^4+1)\right].
\label{2.41}
\end{eqnarray}
\subsubsection{Bound state spinor}\label{sect3d3}
The 4 relevant parameters are
\begin{eqnarray}
e_{12} & = & \frac{2 (Z_1Z_2-\eta^2)(\eta^2 Z_2+Z_1) (\eta^4 Z_1^2Z_2^4 + \eta^4 Z_1^2 - 2 Z_2^2)}
{(Z_2^2+1)\hat{d}}e_1^I
\nonumber \\
e_{112} & = & \frac{2 Z_1 (2 Z_1^2 Z_2^2 - \eta^4 - \eta^4 Z_2^4)}{(Z_2^2+1)(Z_1-\eta^2 Z_2)(\eta^2 + Z_1 Z_2)}e_1^I
\nonumber \\
f_{12} & = & - \frac{2\eta Z_1 (Z_1Z_2-\eta^2)(\eta^2 Z_2+Z_1) (2 \eta^4 Z_1^2Z_2^2 - 1 -  Z_2^4)}
{(Z_2^2+1)\hat{d}}e_1^I
\nonumber \\
f_{112} & = & - \frac{2 \eta ( Z_1^2 Z_2^4 + Z_1^2 - 2 \eta^4 Z_2^2)}{(Z_2^2+1)(Z_1-\eta^2 Z_2)(\eta^2 + Z_1 Z_2)}e_1^I
\label{2.42}
\end{eqnarray}

We reiterate that most of these expressions would look much more complicated if expressed in terms of conventional variables
($k,y_1,y_2,v$). Only the potential $S$ and the time delay factor $\delta$ are fairly simple in either representation. Since these are
also the most interesting quantities, we have collected the results for $S$ and $\delta$ using conventional variables in the appendix
for the convenience of the reader.

\subsection{Self-consistency and fermion density}\label{sect3e}

A crucial step in the TDHF approach still missing so far is the proof of self-consistency. We have to verify that the scalar condensate computed 
from our spinors reproduces the mean field $S$. We first evaluate the scalar density from the above spinors. The result for positive energy
discrete states is
\begin{eqnarray}
(\bar{\psi} \psi)^{(1)} & = & \frac{Z_1^2-1}{i Z_1} \frac{U_1\left(1+ \chi_1 U_2 + \delta U_2^2 \right)}{\cal D},
\nonumber \\
(\bar{\psi} \psi)^{(2)} & = & \frac{Z_2^2-1}{iZ_2} \frac{U_2 \left(\delta + \chi_2 U_1 + U_1^2 \right)}{\cal D},
\label{2.43}
\end{eqnarray}
with 
\begin{eqnarray}
\chi_1 & = & \frac{2 Z_2 \left[ 2 \eta^4 Z_2^2(Z_1^4+1) - (\eta^8+1) Z_1^2(Z_2^4+1) \right]}{(Z_2^2+1)\hat{d}},
\nonumber \\
\chi_2 & = & \frac{2 Z_1 \left[  2 \eta^4 Z_1^2(Z_2^4+1) -  (\eta^8+1) Z_2^2(Z_1^4+1) \right]}{(Z_1^2+1)\hat{d}}.
\label{2.44}
\end{eqnarray}
The negative energy states yield the opposite sign. In the case of the continuum states, we decompose the scalar density
into 2 terms, following the same strategy as in the single baryon case, 
\begin{eqnarray}
\bar{\psi}_{\zeta}\psi_{\zeta}  &  = &  (\bar{\psi}_{\zeta}\psi_{\zeta})_1   + (\bar{\psi}_{\zeta}\psi_{\zeta})_2,
\nonumber \\
(\bar{\psi}_{\zeta}\psi_{\zeta})_1 & = & - \frac{2\zeta}{\zeta^2+1} S ,
\nonumber \\
(\bar{\psi}_{\zeta}\psi_{\zeta})_2   & = &  {\cal F}_{\zeta}^{(1)} (\bar{\psi} \psi)^{(1)} + {\cal F}_{\zeta}^{(2)} (\bar{\psi} \psi)^{(2)} .
\label{2.45}
\end{eqnarray}
The coefficients ${\cal F}_{\zeta}^{(1,2)}$ are ($x,t$)-independent.
Since 3 independent functions of ($x,t$) are needed in order to represent the scalar condensate of an arbitrary single particle state,
we are dealing here with a type III solution of the TDHF problem.
Using the boosted spectral parameters $Z_I, Z_{II}$ introduced in 
(\ref{2.19},\ref{2.21}), the ${\cal F}_{\zeta}^{(1,2)}$ can be represented as
\begin{eqnarray}
{\cal F}_{\zeta}^{(1)} & = & \frac{4i\zeta Z_I^2  (Z_1^4-1)}{(\zeta^2+1)(Z_I^2 - Z_1^2)(Z_I^2 Z_1^2 -1)},
\nonumber \\
{\cal F}_{\zeta}^{(2)} & = & \frac{4i \zeta Z_{II}^2 (Z_2^4-1)}{(\zeta^2+1)(Z_{II}^2-Z_2^2)(Z_{II}^2Z_2^2-1)}.
\label{2.46}
\end{eqnarray}
Integrating the first term in (\ref{2.45}) over $d\zeta$ gives self-consistency, just like in the case of the single baryon.
The integration over the 2nd term yields
\begin{equation}
\int_0^{\infty} \frac{d\zeta}{2\pi} \frac{\zeta^2 +1}{2\zeta^2} (\bar{\psi}_{\zeta} \psi_{\zeta} )_2 = - \frac{i}{2\pi} \left[ (\bar{\psi}\psi)^{(1)} \ln Z_1^4 
+(\bar{\psi} \psi)^{(2)} \ln Z_2^4 \right].
\label{2.47}
\end{equation}
It cancels the contribution from the discrete states provided that the conditions
\begin{equation}
2\pi i (\nu_{i,+} - \nu_{i,-}) + \ln Z_i^4  = 0, \quad (i=1,2)
\label{2.48}
\end{equation}  
hold. These conditions are identical to the result for the single baryon, Eq.~(\ref{1.36}), confirming the self-consistency of 
the scattering solution. 

So far we have only dealt with the massless GN model. As discussed at the end of Sec.~\ref{sect2b}, the single baryon solution also solves 
the massive GN model, with a modified relationship between the parameter $y$ and the occupation fractions $\nu_{\pm}$. As we have seen, 
a prerequisite for this to happen is the proportionality 
\begin{equation}
\langle \bar{\psi}\psi \rangle + \frac{S}{\pi} \ln \Lambda \sim (1-S).
\label{2.49}
\end{equation}
In the baryon-baryon scattering case, this proportionality does not hold (no matter how one chooses the parameters), so that our ansatz does
not lead to a solution of the massive GN model. Presumably, this reflects the fact that the massive GN model is not integrable, so that
inelastic processes will contribute to baryon-baryon scattering if one switches on the bare mass (for a recent discussion of this issue, see 
Ref.~\cite{L22}). 

The fermion density for the bound states can be computed from the spinors,
\begin{equation}
\rho^{(i)} = (\psi^{\dagger}\psi)^{(i)}.
\label{2.50}
\end{equation}
The analytical result is not very instructive and will not be given here (unlike in the single baryon case, it is not proportional to $S^2-1$). 
For the negative energy continuum states we find the subtracted density
\begin{equation}
\psi^{\dagger}_{\zeta} \psi_{\zeta}-1  = - {\cal G}_{\zeta}^{(1)} \rho^{(1)} - {\cal G}_{\zeta}^{(2)} \rho^{(2)},
\label{2.51}
\end{equation}
where the $\rho^{(i)}$ are the densities of the discrete states. The ${\cal G}_{\zeta}^{(i)}$ are ($x,t$)-independent coefficients
satisfying
\begin{equation}
\int_0^{\infty} \frac{d\zeta}{2\pi} \frac{\zeta^2+1}{2\zeta^2} {\cal G}^{(i)}_{\zeta} = 1,  \qquad (i=1,2).
\label{2.52}
\end{equation}
Hence the situation is again similar to the single baryon case. The total (subtracted) density can be computed from the bound state 
densities alone,
\begin{equation}
\rho = \sum_{i=1}^2 (\nu_{i,+} + \nu_{i,-} -1) \rho^{(i)} ,
\label{2.53}
\end{equation}
and the total fermion number is just the sum of the fermion numbers of both baryons. For ground state baryons ($\nu_{i,-}=1$), the induced
fermion density in the Dirac sea is cancelled exactly against the negative energy bound state contributions.
Since the spinors have been given above and the analytical result for the density apparently cannot be simplified very much (unlike the scalar densities),
we refrain from writing it down.

\subsection{General Lorentz frame}\label{sect3f}

We have been working in the Lorentz frame where the baryons move with velocities $\pm v$. Suppose we are interested in a general frame
where the velocities are $v_1,v_2$. Since the Dirac-TDHF approach is covariant, we can transform the scalar potential and the spinors by a
Lorentz boost. Consider the scalar potential $S$ first. The velocity dependence enters in the exponentials $U_1,U_2$, Eqs. (\ref{2.13}), in the 
time delay parameter $\delta$, Eq.~(\ref{2.35}), and in the coefficients $a_{12},b_{12}$, Eqs.~(\ref{2.37}). We evidently have to replace $U_{1,2}$ 
by
\begin{equation}
U_i = \exp \left\{ y_i (\eta_i^{-1} \bar{z}+ \eta_i z)\right\}, \quad \eta_i = e^{\xi_i} = \sqrt{\frac{1+v_i}{1-v_i}}.
\label{2.54}
\end{equation}
All coefficients entering $S$ must be Lorentz scalars. Denoting the baryon 2-velocities by
\begin{equation}
u_i = \gamma_i  \left( \begin{array}{c} 1 \\ v_i \end{array} \right),
\label{2.55}
\end{equation} 
the only non-trivial Lorentz scalar available is 
\begin{equation}
u_1 u_2 = \gamma_1 \gamma_2 (1-v_1v_2).
\label{2.56}
\end{equation}
In the frame where $v_1=-v_2=v$, it reduces to
\begin{equation}
u_1u_2 = \frac{1+v^2}{1-v^2} .
\label{2.57}
\end{equation}
We can then ``covariantize" $S$ by replacing  
\begin{equation}
v^2 \to \frac{u_1u_2-1}{u_1u_2+1} = \frac{\gamma_1 \gamma_2(1-v_1v_2)-1}{\gamma_1\gamma_2(1-v_1v_2)+1}= v_{12}^2
\label{2.58}
\end{equation}
where $v_{12}$ is the (relativistic) relative velocity   
\begin{equation}
v_{12} = \frac{1-v_1v_2 - \sqrt{(1-v_1^2)(1-v_2^2)}}{v_1-v_2}.
\label{2.59}
\end{equation}
This rule is applicable to $S$ or $\delta$ written in conventional variables, as given in the appendix. 
As expected, the 
transformation to another Lorentz frame is more elegant in lightcone variables, where it reduces to the simple substitution rule
\begin{equation}
\eta \to \sqrt{\frac{\eta_1}{\eta_2}}.
\label{2.60}
\end{equation}
This transforms the time delay factor $\delta$ into the symmetric expression
\begin{equation}
\delta = \frac{(\eta_2 Z_1+ \eta_1 Z_2)(\eta_1 Z_1 + \eta_2 Z_2)(\eta_1 Z_1 Z_2-\eta_2)(\eta_2 Z_1Z_2-\eta_1)}
{(\eta_2 Z_1- \eta_1 Z_2)(\eta_1 Z_1 - \eta_2 Z_2)(\eta_1 Z_1 Z_2+\eta_2)(\eta_2 Z_1Z_2+\eta_1)}.
\label{2.61}
\end{equation}
Likewise, the complexity of $a_{12},b_{12}$, Eqs.~(\ref{2.37}), does not increase if one replaces $\eta$ by $\sqrt{\eta_1/\eta_2}$.

Finally, we remark that an even simpler form for $\delta$ can be obtained by going back to Eq.~(\ref{2.33}) for the bound state transmission 
amplitude. Let us transform $T_1^{II}$ into an arbitrary frame, using the substitution (\ref{2.60}),
\begin{equation}
T_1^{II} = \frac{(\zeta_1^*+ \zeta_2^*)(\zeta_1^*-\zeta_2)}{(\zeta_1^*- \zeta_2^*)(\zeta_1^*+\zeta_2)}
\label{2.62}
\end{equation}
with $\zeta_1 = \eta_1 Z_1, \zeta_2 = \eta_2 Z_2$. This leads to the most compact expression for $\delta$ in an arbitrary Lorentz frame which we
could find,
\begin{equation}
\delta = \left| \frac{(\zeta_1+\zeta_2)(\zeta_1-\zeta_2^*)}{(\zeta_1-\zeta_2)(\zeta_1+\zeta_2^*)}\right|^2.
\label{2.63}
\end{equation}
The standard observable in 1d scattering of solitons is the time delay, closely related to $\delta$. Baryons $I$ and $II$ can be characterized
by the following exponentials in the incoming and outgoing channels,
\begin{eqnarray}
U_1^{\rm in} & = &  \exp \left\{ 2 y_1 \gamma_1(x-v_1t) \right\}, \quad U_1^{\rm out} = \delta^{-1} U_1^{\rm in},
\nonumber \\
U_2^{\rm in} & = &  \exp \left\{ 2 y_2 \gamma_2(x-v_2t) \right\}, \quad U_2^{\rm out} = \delta U_1^{\rm in},
\label{2.63a}
\end{eqnarray} 
The time delays $\Delta t_1, \Delta t_2$ are introduced via
\begin{equation}
U_i^{\rm out}(x,t) = U_i^{\rm in}(x,t - \Delta t_i), \quad (i=1,2)
\label{2.63b}
\end{equation}
and given by
\begin{equation}
\Delta t_1 = -  \frac{\ln \delta}{2 y_1 \gamma_1 v_1}, \quad \Delta t_2 =  \frac{\ln \delta}{2 y_2 \gamma_2 v_2}.
\label{2.63c}
\end{equation}

So far, we have only discussed the transformation of $S$ and $\delta$ into a general Lorentz frame. The spinors can also easily be boosted
in the lightcone approach, since the boost matrix $e^{\xi \gamma_5/2}$ is diagonal. By contrast, the corresponding formulae in normal
coordinates become exceedingly complicated.

\section{Limiting cases and illustrative examples}\label{sect4}

We have presented above the general solution of baryon-baryon scattering in the large $N$ limit of the GN model. In certain regions of 
parameter space, our solution can be compared to previous works. This gives us the opportunity to crosscheck our results against the 
existing literature. The special cases we are aware of are the static limit ($v_1=v_2=0$), the non-relativistic limit ($v_1,v_2,y_1,y_2 \ll 1$) and
the kink limit ($y_1, y_2\to 1$), i.e., the ultrarelativistic limit for the internal motion. We will also illustrate the full results
with the help of some examples and make contact with the scattering of polarons and solitons in trans-polyacetylene in the present section.

\subsection{Static limit}\label{sect4a}

All static, transparent potentials of the 1d Schr\"odinger equation have been constructed long ago by Kay and Moses \cite{L7}. These 
Schr\"odinger potentials also enter in the construction of a complete set of static, transparent potentials of the Dirac equation, which are at the 
same time self-consistent potentials of the GN model \cite{L6,L8}. Physically, they correspond to marginally bound $n$-baryon states whose 
mass is the sum of the constituent masses, independent of their separation. The $n=2$ case can be related to the static limit
($v_1=v_2=0$) of baryon-baryon scattering. In the static case, one starts from the linear system of equations
\begin{equation}
\sum_{j=1}^n A_{ij} \psi_j = \lambda_i, \qquad (i=1,...,n)
\label{3.1}
\end{equation}
where
\begin{equation}
A_{ij} = \delta_{ij}+ \frac{\lambda_i \lambda_j}{\kappa_i + \kappa_j}, \quad \lambda_i = c_i e^{-\kappa_ix}.
\label{3.2}
\end{equation}
The Dirac-HF potential is then given by \cite{L6}
\begin{equation}
S=1- \partial_x \ln \left( 1- \sum_{i=1}^n \frac{\lambda_i \psi_i}{1+ \kappa_i}\right).
\label{3.3}
\end{equation}
Comparing our results for $v=0$  with the static  $n=2$  solution, we get perfect agreement if we employ the following dictionary,
\begin{eqnarray}
y_1 & = &  \kappa_1, \quad y_2 = \kappa_2,
\nonumber \\
U_1 & = & A_1 e^{2 y_1 x}, \quad U_2 = A_2 e^{2 y_2 x},
\nonumber \\
A_1 & = & \frac{2 y_1 (1+y_1)}{c_1^2 w_1} \delta, \quad \delta= \left( \frac{y_1+y_2}{y_1-y_2} \right)^2,
\nonumber \\
A_2 & = & \frac{2 y_2 w_2}{c_2^2 (1-y_2)}.
\label{3.4}
\end{eqnarray}

\subsection{Non-relativistic limit}\label{sect4b}

Nogami and Warke \cite{L23} have constructed transparent potentials of the time-dependent, 1d Schr\"odinger equation, thereby
solving the non-relativistic TDHF problem for particles with $\delta$-interactions. Equivalently, this amounts to finding solutions of the 
multicomponent nonlinear Schr\"odinger equation. We should be able to recover their results in the limit where both the internal motion
of the fermions in the DHN baryons and their external motion as a whole are non-relativistic, i.e., for $y_1,y_2,v_1,v_2 \ll 1$.  
The non-relativistic, time-dependent construction of Ref.~\cite{L23} generalizes the static one as follows: One starts from the linear
system of equations
\begin{equation}
\sum_{\beta=1}^n \frac{u_{\alpha} u_{\beta}^* g_{\beta}}{\kappa_{\alpha}+ \kappa_{\beta}^*} + g_{\alpha} + u_{\alpha} = 0
\label{3.5}
\end{equation}
with
\begin{equation}
u_{\alpha} = \sqrt{A_{\alpha}} \exp \left\{ \kappa_{\alpha}x + i \kappa_{\alpha}^2 t \right\}.
\label{3.6}
\end{equation}
The $A_{\alpha}$ are real, the $\kappa_{\alpha}$ complex parameters. The self-consistent potential can be written as
\begin{equation}
V=2 \partial_x \sum_{\alpha} u_{\alpha}^* g_{\alpha}.
\label{3.7}
\end{equation}
We have evaluated $V$ for $n=2$ and compared it with the non-relativistic limit of our $S$ in a frame where the baryon velocities are $v_1,v_2$.
We treat $y_1,y_2,v_1,v_2$ as small quantities of order $\epsilon$ and keep only the leading order term in ($S-1$), which is of order $\epsilon^2$.
Due to the fact that Nogami and Warke use units where $m=1/2$ (to simplify the Schr\"odinger equation) whereas we use units
where $m=1$, we also have to rescale time $t$ by a factor of 2 and compare their $V$ with our $2(S-1)$. The two potentials then agree for the
following choice of parameters,
\begin{eqnarray}
\kappa_1 & = &  y_1+i v_1, \quad \kappa_2 = y_2 + i v_2,
\nonumber \\
A_1 & = &  2 y_1, \quad A_2 = 2 y_2 \delta,
\nonumber \\
\delta & = &  \frac{(y_1+y_2)^2+(v_1-v_2)^2}{(y_1-y_2)^2+(v_1-v_2)^2}.
\label{3.8}
\end{eqnarray}

\subsection{Kink limit}\label{sect4c}

In the limit $y_1\to 1, y_2\to 1$, the internal structure of the DHN baryon becomes ultrarelativistic. The baryons decouple into well separated
kink and antikink. We should like to compare our result for $S$ in this limit to the result for kink-antikink scattering \cite{L13,L14},
\begin{equation}
S= \frac{v \cosh 2\gamma x - \cosh 2 \gamma vt}{v \cosh 2\gamma x + \cosh 2 \gamma vt}.
\label{3.9}
\end{equation}
Since the DHN baryon becomes infinitely extended in the kink limit, we have to shift our exponentials $U_1,U_2$ in such a way that only the
scattering of the right edge of baryon $I$ (an antikink) and the left edge of baryon $II$ (a kink) survives in the limit $y_1,y_2\to 1$. Working
in the center-of-velocity frame, we first set
\begin{equation}
y_1=y_2=1-\epsilon.
\label{3.10}
\end{equation}
The exponentials $U_1,U_2$ are replaced by
\begin{equation}
U_1 = A \exp \left\{ 2 \gamma(x-vt)\right\}, \quad U_2 = A^{-1} \exp \left\{ 2 \gamma(x+vt)\right\}.
\label{3.11}
\end{equation}
For the choice 
\begin{equation}
A= \frac{\sqrt{\epsilon}}{v^3 \sqrt{2}}
\label{3.12}
\end{equation}
one can then perform the limit $\epsilon \to 0$ without encountering any singularity and $S$ goes over into the negative of (\ref{3.9}). The change of sign occurs because we get 
antikink-kink scattering, whereas (\ref{3.9}) holds for kink-antikink scattering.

\subsection{Examples}\label{sect4d}

\begin{figure}[h]
\begin{center}
\epsfig{file=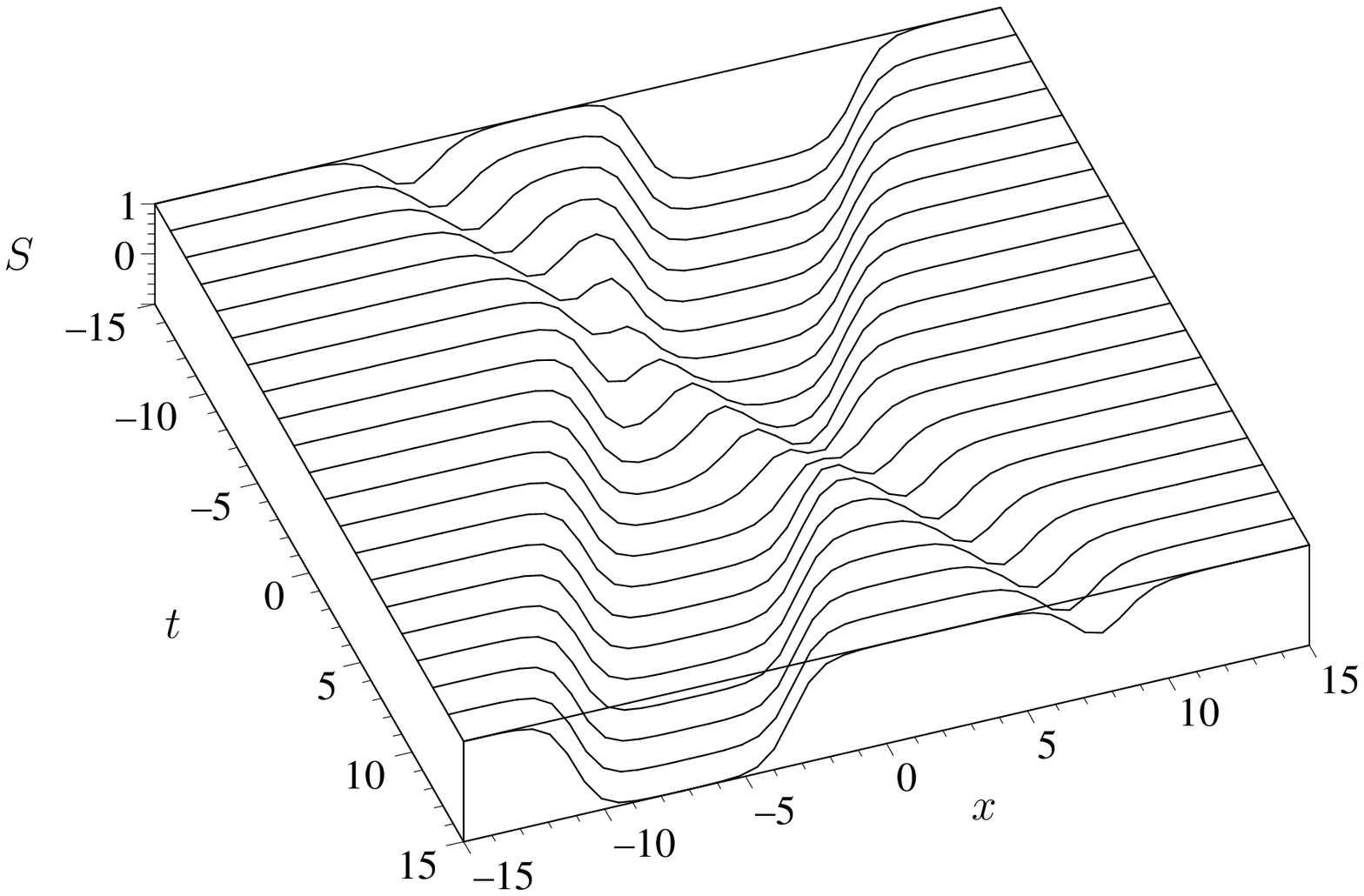,angle=270,width=8cm}
\caption{Scattering of a small ($y_1=0.8$) and a large ($1-y_2=10^{-7}$) DHN baryon with velocities $\pm 0.4$. The self-consistent scalar potential $S$ is shown
for a range of ($x,t$) values in the vicinity of the collision.}
\label{fig1}
\end{center}
\end{figure}
\begin{figure}[h]
\begin{center}
\epsfig{file=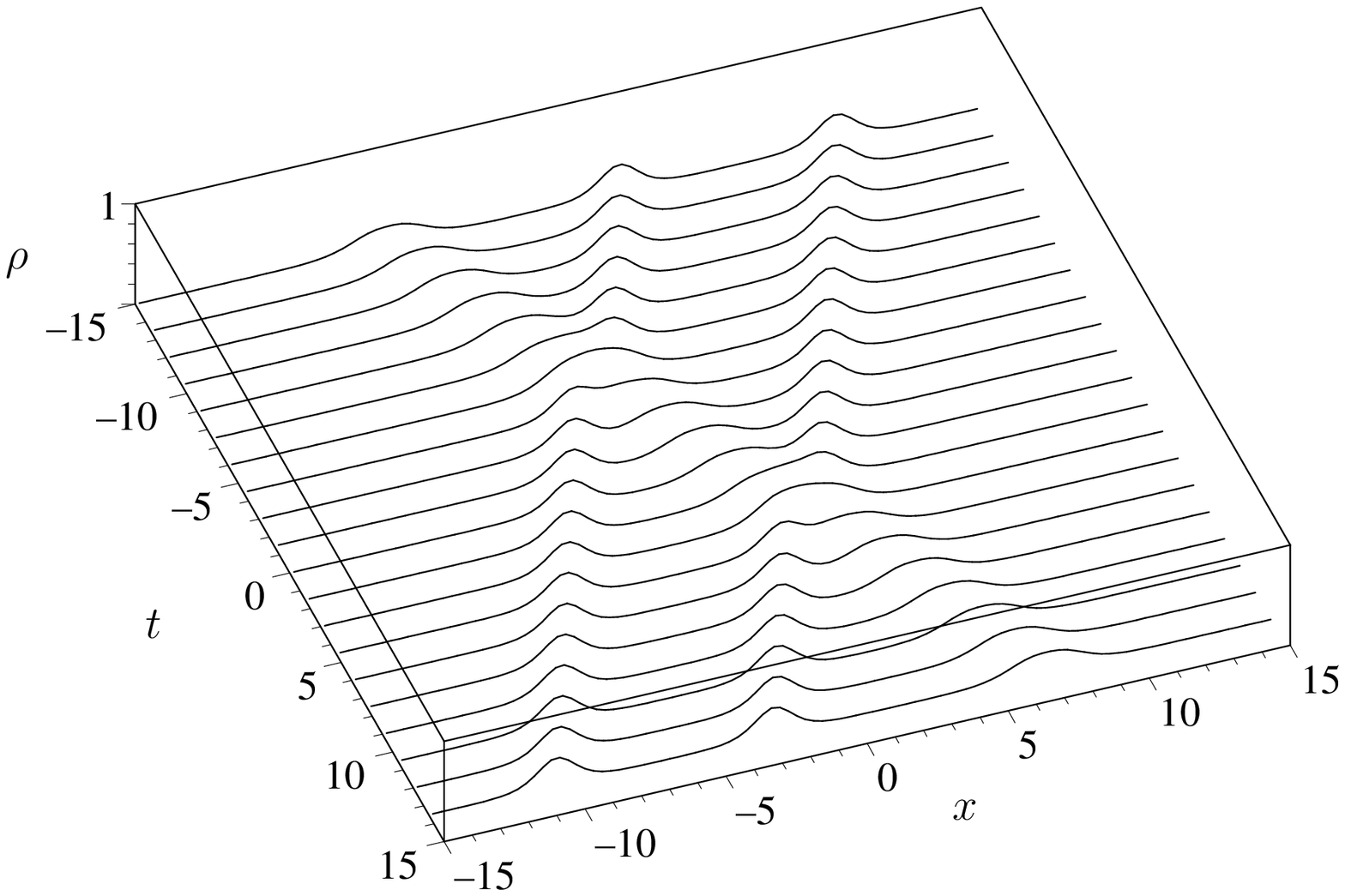,angle=270,width=8cm}
\caption{Same as Fig.~\ref{fig1}, but total fermion density $\rho$ shown. In the large baryon, the density has two peaks near the kink and antikink boundaries of the 
baryon. A negative time delay occurs during the collision.}
\label{fig2}
\end{center}
\end{figure}
In our first example, we consider scattering of a ``small" baryon ($y_1=0.8$) and  a ``large" baryon ($1-y_2=10^{-7}$), with velocities $\pm 0.4$. Fig.~\ref{fig1}
shows the scalar potential as a function of $x$, for different time slices. The baryons cross each other with some (negative) time delay, as can also be seen 
from Fig.~\ref{fig2} where the fermion densities are shown for the same scattering process. This should be contrasted to kink-antikink scattering where
the scalar potentials first approach and then repel each other, the fermions being exchanged during the collision \cite{L13}. It is also interesting to 
watch the small baryon while it is crossing the large one in Fig.~\ref{fig1}. In the initial and final state, the small baryon shows the ususal attractive potential well
in the vacuum value $S=1$, the large baryon exhibits well separated kink and antikinks with $S=-1$, i.e., the other degenerate vacuum, inbetween. 
During the collision, the small baryon moves as a seemingly repulsive potential bump on the $S=-1$ ``floor". This is just the chirally reflected DHN baryon in 
the other vacuum. To emphasize this aspect, we consider in our second example scattering of a small DHN baryon ($y_1=0.8$) on a CCGZ kink ($y_2=1$), with
velocities $\pm 0.2$. In this case of baryon-kink scattering, the formula for $S$ simplifies to the following expression (in conventional coordinates),
\begin{equation}
S= \frac{1+ \frac{2(1-2y_1^2)}{w_1}U_1 - \delta U_2 + U_1^2- \frac{2(1-2 y_1^2)}{w_1}\sqrt{\delta} U_1 U_2 - U_1^2 U_2}{1+ \frac{2}{w_1}U_1 + 
\delta U_2 + U_1^2 + \frac{2}{w_1} \sqrt{\delta} U_1 U_2 + U_1^2 U_2},
\label{3.13}
\end{equation}
with $y_2=1$ and
\begin{equation}
\delta = \left( \frac{1 + v^2 + y_1(1-v^2)}{1 + v^2 -y_1(1-v^2)}\right)^2.
\label{3.14}
\end{equation}
\begin{figure}[h]
\begin{center}
\epsfig{file=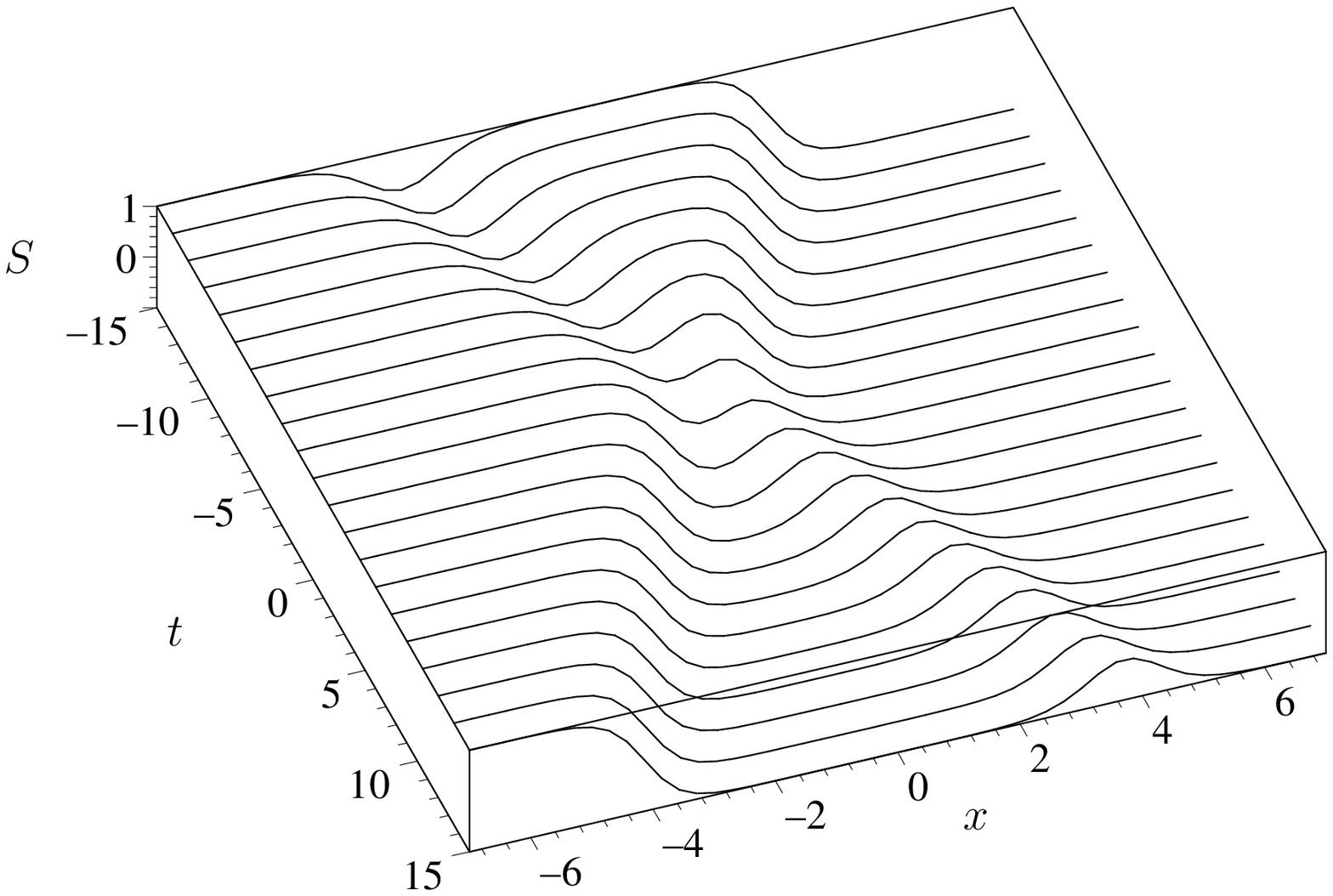,angle=270,width=8cm}
\caption{Scattering of a small DHN baryon ($y=0.8$) on a CCGZ kink ($y=1$), at velocities $\pm 0.2$. Notice the chiral reflection which the small baryon undergoes 
during the collision.}
\label{fig3}
\end{center}
\end{figure}
Comparing the small baryons in the first and last time slice of Fig.~\ref{fig3}, one sees clearly the sign flip of the potential characteristic for the discrete $\gamma_5$ transformation during
the collision.

The last few examples are applications of the GN model to condensed matter physics. As is well known, the Su-Schrieffer-Heeger theory \cite{L24} of trans-polyacetylene
admits a continuum approximation \cite{L25} closely related to the massless GN model. At the mean field level, the correspondence with the large $N$ GN model
is perfect, although $N$ in the condensed matter case is only 2, the number of electron spin components. DHN baryons become polarons, CCGZ kinks
become solitons in conducting polymer language \cite{L26}. The polaron has one electron in the upper bound state, correponding to the DHN parameter $y=\sin \pi/4$.
For polaron-polaron scattering with relative velocity $v$, we thus have to choose $y_1=y_2=1/\sqrt{2}$ and find $\delta=1/v^2$,
\begin{equation}
S= \frac{1+ U_1^2 + \frac{2(1- v^2)^2}{v^2(1+v^2)} U_1 U_2 +  \frac{1}{v^4} U_2^2 + U_1^2 U_2^2}{1+ 2\sqrt{2}U_1 + \frac{2\sqrt{2}}{v^2} U_2
+ U_1^2 + \frac{2(1+6v^2+v^4)}{v^2(1+v^2)}U_1U_2 + \frac{1}{v^4}U_2^2 + 2\sqrt{2}U_1^2 U_2 + \frac{2 \sqrt{2}}{v^2} U_1 U_2^2+ U_1^2U_2^2}.
\label{3.15}
\end{equation}
Polaron-soliton scattering on the other hand corresponds to the parameters $\pm v$ and $y_1=1/\sqrt{2},y_2=1$, where $S$ simplifies to
\begin{equation}
S = \frac{1-\delta U_2+ U_1^2 - U_1^2U_2}{1 + 2 \sqrt{2}U_1 + \delta U_2 + U_1^2 +  2\sqrt{2} \sqrt{\delta} U_1 U_2 + U_1^2 U_2}
\label{3.16}
\end{equation}
with
\begin{equation}
\delta = \left( \frac{(1+v^2)\sqrt{2} + (1-v^2)}{(1+v^2)\sqrt{2}-(1-v^2)}\right)^2.
\label{3.17}
\end{equation}
Finally, soliton-soliton scattering can be inferred from Eq.~(\ref{3.9}) in Sec.~\ref{sect3c}.

\section{Summary and outlook}\label{sect5}

In this paper we have presented a complete description of the scattering of two DHN baryons, each with its own internal structure of localized mean fields and bound 
fermions, in the large $N$ limit of the Gross-Neveu (GN) model. We have used the time-dependent Hartree-Fock (TDHF) method, thereby realizing an explicit example of the picture 
envisioned by Witten in his
seminal paper on baryons in the $1/N$ expansion \cite{L27}.
Static baryons of this model have been known since the early work of DHN, but progress on time-dependent, self-consistent
solutions has been slow until very recently. The only cases which had been understood so far are the non-relativistic limit, where the model
reduces to the multicomponent nonlinear Schr\"odinger equation, and the ultrarelativistic limit of kinks and antikinks. 
Here we have presented the most general solution, valid for 2 DHN baryons with arbitrary fermion numbers and velocities.
This contains in particular scattering of polarons and solitons in the continuum limit of conducting polymers like trans-polyacetylene. 
The solution displays characteristic features expected from an integrable model: Purely elastic scattering, vanishing reflection coefficients, baryons crossing each
other with a time delay. The complete time evolution, including the non-asymptotic region, is available in closed analytical form through the TDHF spinors.
The calculation is significantly more involved than in the kink-antikink case, but remains manageable owing to the systematic use of lightcone coordinates
and variables with their simple transformation properties under Lorentz boosts.  

More interesting than the specific results for baryon-baryon scattering is perhaps the method used to solve this problem. Since the scattering solution
is of type III, the TDHF problem is mathematically equivalent to solving 3 coupled, nonlinear partial differential equations. It does not seem very likely that one
can guess the analytical solution of such a complicated problem. Nevertheless, the method of ansatz has proven very effective. The basic idea was
to use a simultaneous ansatz for the TDHF potential and the single particle spinors. This ansatz is based on very few heuristic rules: The scalar potential
is assumed to be a rational function of 2 exponentials, one for each baryon. The degree of the polynomials in the numerator and denominator is taken from the
product of two independent single baryon potentials. The coefficients are determined to a large extent by the known baryons in the incoming and outgoing states.
The spinors are also assumed to be rational functions of exponentials times a plane wave factor (for continuum states) or its analytic continuation to complex
spectral parameter (for bound states). The denominators are taken to be the same in $S$ and all the spinors, the numerators have different coefficients fixed largely by
asymptotics. When the dust has settled down, only a handful of coefficients remain to be determined by solving the Dirac equation, which is now turned into a simple
algebraic problem. Having found a solution of the Dirac equation, we still have to check self-consistency of the whole approach. This was successful for the
massless GN model but not for the massive one, although single DHN baryons solve both of these models.

Our method can be generalized in a natural way to other, even more complicated, baryon scattering problems. We have already used it to reproduce the DHN breather
and to solve a dynamical 3-baryon problem. This will be reported elsewhere. One could also consider the scattering of a DHN baryon from a lump of finite-density matter,
 represented as an array of kinks and antikinks, or even the scattering of two such lumps. The question then arises whether one can find the general $multi$-baryon solution
of the TDHF equation in closed form, similar to what has already been achieved in the static limit, in the non-relativistic limit or in the kink limit of the
GN model. The two-baryon solution discussed here has many encouraging features which suggest that such a generalization might actually exist.
If one can find it, it will be interesting to see whether it coincides with some other solved problem in nonlinear mathematical physics, like in the case of the 
sinh-Gordon theory, or whether it leads to a new class of exactly solvable problems. We hope that having worked out the 2-baryon problem in full detail,
and having cast the solution into the simplest form, will help to answer this question in the future.

\section*{Acknowledgement}

The work of G.~D. has been supported by the US DOE under Grant DE-FG02-92ER40716, and the work of C. F. and M. T. has been supported by the DFG under grant TH 842/1-1. 
One of us (M. T.) thanks the Department of Physics of the University of Connecticut, Storrs, for the hospitality extended to him during a stay where part of this work was done.

\section*{APPENDIX: SCALAR POTENTIAL IN CONVENTIONAL VARIABLES}
The self-consistent scalar potential for baryon-baryon scattering has the general form introduced in Eqs.~(\ref{2.1})-(\ref{2.4}). Here we list
all the parameters in conventional variables, complementing the lightcone variables used in the main text. Parameters determined 
asymptotically by one-baryon data,
\begin{eqnarray}
a_1 & = & \frac{2(1-2y_1^2)}{w_1}, \quad a_2 = \frac{2(1-2 y_2^2)}{w_2}\delta, 
\nonumber \\
a_{112} & = & \frac{2(1-2 y_2^2)}{w_2}, \quad a_{122} = \frac{2(1-2y_1^2)}{w_1} \delta,
\nonumber \\
b_1 & = & \frac{2}{w_1}, \quad b_2 = \frac{2}{w_2}\delta,
\nonumber \\
b_{112} & = & \frac{2}{w_2}, \quad b_{122} = \frac{2}{w_1}\delta,
\nonumber \\
a_{11} & = &  a_{1122} = b_{11} = b_{1122} = 1,
\nonumber \\
a_{22} & = &  b_{22} = \delta^2,
\label{A1}
\end{eqnarray}
with $w_i = \sqrt{1-y_i^2}$ and 
\begin{equation}
\delta  = \frac{n}{d} = \frac{4 v^2 +(1-v^2)^2 (y_1^2+y_2^2) + 2 (1-v^4) y_1y_2}{4 v^2 +(1-v^2)^2 (y_1^2+y_2^2) - 2(1-v^4) y_1y_2}.
\label{A2}
\end{equation}
Parameters determined via the Dirac equation,
\begin{eqnarray}
a_{12} & = &  \frac{4 [4 v^2+ (1-10v^2+v^4)(y_1^2+y_2^2)-2(1-v^2)^2(y_1^4+y_2^4)+ 2(1+6v^2+v^4)y_1^2y_2^2]}
{w_1 w_2 d},
\nonumber \\
b_{12} & = & \frac{4 [4 v^2 + (1-v^2)^2(y_1^2+y_2^2-2 y_1^2y_2^2) ]}
{w_1 w_2 d},
\label{A3}
\end{eqnarray}
with $d$ from Eq.~(\ref{A2}).

In a Lorentz frame where the baryons move with velocities $v_1,v_2$, replace $U_1,U_2$ in Eqs.~(\ref{2.4}) by
\begin{equation}
U_i = \exp \left\{ 2 y_i \gamma_i (x-v_i t) \right) , \quad \gamma_i = \frac{1}{\sqrt{1-v_i^2}} \quad(i=1,2),
\label{A4}
\end{equation}
and replace $v$ by the relative velocity $v_{12}$, Eq.~(\ref{2.59}), in $\delta, a_{12}, b_{12}$.


\begin{thebibliography}{99}
\bibitem{L1}
D. J. Gross and A. Neveu, Phys. Rev. D {\bf 10}, 3235 (1974).
\bibitem{L2}
D. J. Gross and F. Wilczek, Phys. Rev. Lett. {\bf 30}, 1343 (1973).
\bibitem{L3}
H. D. Politzer, Phys. Rev. Lett. {\bf 30}, 1346 (1973).
\bibitem{L4}
R. F. Dashen, B. Hasslacher and A. Neveu, Phys. Rev. D {\bf 12}, 2443 (1975).
\bibitem{JR}
R. Jackiw, C. Rebbi,  Phys. Rev. D {\bf 13}, 3398 (1976).
\bibitem{L5}
D. J. Gross, in Les Houches 1975, Proceedings, {\it Methods in Field Theory}, Amsterdam (1976), p. 141-250.
\bibitem{L6}
J. Feinberg, Ann. Phys. (N.Y.) {\bf 309}, 166 (2004).
\bibitem{L7}
I. Kay and H. E. Moses, J. Appl. Phys. {\bf 27}, 1503 (1956).
\bibitem{L8}
F. M. Toyama, Y. Nogami, Z. Zhao, Phys. Rev. A {\bf 47}, 897 (1993).
\bibitem{L9}
M. Thies, Phys. Rev. D {\bf 69}, 067703 (2004).
\bibitem{L10}
U. Wolff, Phys. Lett. B {\bf 157}, 303 (1985).
\bibitem{L11}
O. Schnetz, M. Thies, K. Urlichs, Ann. Phys. {\bf 314}, 425 (2004). 
\bibitem{L27}
E. Witten, Nucl. Phys. B {\bf 160}, 57 (1979).
\bibitem{L12}
W. Brendel and M. Thies, Phys. Rev. D {\bf 81}, 085002 (2010).
\bibitem{L13}
A. Klotzek and M. Thies, J. Phys. A {\bf 43}, 375401 (2010).
\bibitem{L14}
A. Neveu and N. Papanicolaou, Commun. Math. Phys. {\bf 58}, 31 (1978).
\bibitem{L15}
C. Fitzner and M. Thies, Phys. Rev. D {\bf 83}, 085001 (2011).
\bibitem{L16}
G. Basar and G. V. Dunne, JHEP{\bf 1101}, 127 (2011).
\bibitem{L16a}
A. Jevicki and K. Jin, Int. J. Mod. Phys. A {\bf 23}, 2289 (2008).
\bibitem{L17}
A. B. Zamolodchikov and Al. B. Zamolodchikov, Phys. Lett. B {\bf 72}, 481 (1978).
\bibitem{L18}
R. Shankar and E. Witten, Nucl. Phys. B {\bf 141}, 349 (1978).
\bibitem{L19}
M. Karowski and H. J. Thun, Nucl. Phys. B {\bf 190}, 61 (1981).
\bibitem{L20}
M. Thies and K. Urlichs, Phys. Rev. D {\bf 71}, 105008 (2005).
\bibitem{L21}
J. Feinberg and S. Hillel, Phys. Rev. D {\bf 71}, 105009 (2005).
\bibitem{L22}
D. Fernandez-Fraile, Phys. Rev. D {\bf 83}, 065001 (2011).
\bibitem{L23}
Y. Nogami and C. S. Warke, Phys. Lett. A {\bf 59}, 251 (1976).
\bibitem{L24}
W.-P. Su, J. R. Schrieffer, A. Heeger, Phys. Rev. Lett. {\bf 42}, 1698 (1979).
\bibitem{L25}
H. Takayama, Y. R. Lin-Liu, K. Maki, Phys. Rev. B {\bf 21}, 2388 (1980).
\bibitem{L26}
D. K. Campbell and A. R. Bishop, Phys. Rev. B {\bf 24}, 4859 (1981).
\end{thebibliography}
\end{document}